\documentclass[10pt,aps,prb,twocolumn,superscriptaddress,showpacs,longbibliography,floatfix]{revtex4-2}

\usepackage{amssymb}
\usepackage{graphicx}
\usepackage{amsmath}
\usepackage{mathtools}
\usepackage[export]{adjustbox}
\usepackage{epsfig}
\usepackage{times}
\usepackage{color}
\usepackage{subfigure}
\usepackage{setspace}
\usepackage{natbib}
\usepackage[normalem]{ulem}
\usepackage{cancel}
\usepackage{soul}
\usepackage{physics}
\usepackage[left]{lineno}

\usepackage[pdftex]{hyperref}
\usepackage{ulem} 
\hypersetup{colorlinks = true, urlcolor = blue, linkcolor = blue, citecolor = blue}

\usepackage{xcolor}
\usepackage{xparse,xcoffins}
\ExplSyntaxOn
\NewCoffin\imagecoffin
\NewCoffin\labelcoffin
\keys_define:nn { miguel/label }
 {
  label   .tl_set:N = \l_miguel_label_tl,
  labelbox .bool_set:N = \l_miguel_label_box_bool,
  labelbox .default:n = true,
  fontsize .tl_set:N = \l_miguel_label_size_tl,
  fontsize .initial:n = \footnotesize,
  pos .choice:,
  pos/nw .code:n = \tl_set:Nn \l_miguel_label_pos_tl { left,up },
  pos/ne .code:n = \tl_set:Nn \l_miguel_label_pos_tl { right,up },
  pos/sw .code:n = \tl_set:Nn \l_miguel_label_pos_tl { left,down },
  pos/se .code:n = \tl_set:Nn \l_miguel_label_pos_tl { right,down },
  pos/n .code:n = \tl_set:Nn \l_miguel_label_pos_tl { hc,up },
  pos/w .code:n = \tl_set:Nn \l_miguel_label_pos_tl { left,vc },
  pos/s .code:n = \tl_set:Nn \l_miguel_label_pos_tl { hc,down },
  pos/e .code:n = \tl_set:Nn \l_miguel_label_pos_tl { right,vc },
  pos .initial:n = nw,
  unknown .code:n   = \clist_put_right:Nx \l_miguel_label_clist
                       { \l_keys_key_tl = \exp_not:n { #1 } }
 }
\clist_new:N \l_miguel_label_clist
\box_new:N \l_miguel_label_box
\box_new:N \l_miguel_label_image_box
\NewDocumentCommand{\xincludegraphics}{O{}m}
 {
  \group_begin:
  \tl_clear:N \l_miguel_label_tl
  \clist_clear:N \l_miguel_label_clist
  \keys_set:nn { miguel/label } { #1 }
  \tl_if_empty:NTF \l_miguel_label_tl
   {
    \miguel_includegraphics:Vn \l_miguel_label_clist { #2 }
   }
   {
    \SetHorizontalCoffin\imagecoffin
     {
      \miguel_includegraphics:Vn \l_miguel_label_clist { #2 }
     }
    \SetHorizontalCoffin\labelcoffin
     {
      \raisebox{\depth}
       {
        \bool_if:NTF \l_miguel_label_box_bool
         { \fcolorbox{white}{white}{\l_miguel_label_size_tl\l_miguel_label_tl} }
         { \l_miguel_label_size_tl\l_miguel_label_tl }
       }
     }
    \SetVerticalPole\imagecoffin{left}{3pt+\CoffinWidth\labelcoffin/2}
    \SetVerticalPole\imagecoffin{right}{\Width-3pt-\CoffinWidth\labelcoffin/2}
    \SetHorizontalPole\imagecoffin{up}{\Height-3pt-\CoffinHeight\labelcoffin/2}
    \SetHorizontalPole\imagecoffin{down}{3pt+\CoffinHeight\labelcoffin/2}
    \use:x{\JoinCoffins\imagecoffin[\l_miguel_label_pos_tl]\labelcoffin[vc,hc]} 
    \TypesetCoffin\imagecoffin
   }
   \group_end:
 }
\NewDocumentCommand{\setlabel}{m}
 {
  \keys_set:nn { miguel/label } { #1 }
 }
\cs_new_protected:Nn \miguel_includegraphics:nn
 {
  \includegraphics[#1]{#2}
 }
\cs_generate_variant:Nn \miguel_includegraphics:nn { V }
\ExplSyntaxOff

\begin{document}

\title{Modeling Quantum Geometry for Fractional Chern Insulators with unsupervised learning}
\author{Ang-Kun Wu$^{*}$}
\email{angkunwu@gmail.com}
\affiliation{Department of Physics and Astronomy, University of Tennessee, Knoxville, Knoxville, Tennessee, 37996, USA}
\author{Louis Primeau}
\affiliation{Department of Physics and Astronomy, University of Tennessee, Knoxville, Knoxville, Tennessee, 37996, USA}
\author{Jingtao Zhang}
\affiliation{Google, Mountain View, CA, 94043, USA}
\author{Kai Sun}
\affiliation{Department of Physics, University of Michigan, Ann Arbor, Michigan, 48109, USA}
\author{Yang Zhang}
\affiliation{Department of Physics and Astronomy, University of Tennessee, Knoxville, Knoxville, Tennessee, 37996, USA}
\author{Shi-Zeng Lin$^{*}$}
\email{szl@lanl.gov}
\affiliation{Theoretical Division, T-4 and CNLS, Los Alamos National Laboratory (LANL),
Los Alamos, New Mexico 87545, USA}
\affiliation{Center for Integrated Nanotechnology, Los Alamos National Laboratory (LANL),
Los Alamos, New Mexico 87545, USA}

\begin{abstract}
Fractional Chern insulators (FCIs) in moir\'e materials present a unique platform for exploring strongly correlated topological phases beyond the paradigm of ideal quantum geometry. While analytical approaches to FCIs and fractional quantum Hall states (FQHS) often rely on idealized Bloch wavefunctions, realistic moir\'e models lack direct tunability of quantum metric and Berry curvature, limiting theoretical and numerical exploration. Here, we introduce an unsupervised machine learning framework to model interacting Hamiltonians directly through the distribution of single-particle form factors. 
Using a variational autoencoder (VAE), we show that unsupervised learning can not only distinguish FCI and non-FCI states, but also generate new form factors with distinct topological character, not present in the training set.
This latent space enables the generation and interpolation of form factors for topological flatbands with Chern number $|C|=1$, enabling the discovery of unobserved many-body states such as charge density waves. 
Principal component analysis (PCA) further reveals that the dominant patterns in the form factors—reflecting correlations across the Brillouin zone—can be decomposed into components with approximately quantized Chern numbers, providing new insights into the global and topological structure of quantum geometry. 
Our results highlight the ability of machine learning to generalize and model topological quantum systems, paving the way for the inverse design of form factors with tailored quantum geometry and many-body phases in flatband materials.
\end{abstract}

\date{\today}

\maketitle



%
%

\section{Introduction}

Fractional Chern insulators (FCIs) have emerged as a promising platform for realizing exotic topological phases of matter in lattice systems without external magnetic fields~\cite{Bergholtz2013, Parameswaran2013, PhysRevLett.106.236804,neupert2015fractional,LIU2024515,PhysRevResearch.5.L032022,PhysRevLett.132.096602,PhysRevLett.133.156504}. Analogous to fractional quantum Hall states (FQHS)~\cite{PhysRevLett.48.1559, PhysRevLett.50.1395,RevModPhys.71.875,PhysRevLett.90.016801,jain2007composite}, FCIs arise from the interplay between strong electronic correlations and nontrivial band topology, typically characterized by a nonzero Chern number. Unlike continuum Landau level systems, FCIs are realized in flatbands engineered by lattice effects, such as those found in moiré materials~\cite{cai2023signatures, park2023observation, PhysRevX.13.031037,zeng2023thermodynamic,kang2024evidence,lu2024fractional,xie2024even}, where the quantum geometry of Bloch wavefunctions plays a crucial role in stabilizing fractionalized phases~\cite{PhysRevB.90.165139,PhysRevLett.127.246403}.

The study of fractional quantum Hall states (FQHS) \cite{RevModPhys.71.S298} and fractional Chern insulators (FCIs) \cite{PhysRevX.1.021014} centers on the physics of strongly correlated, topologically nontrivial flatbands. In these systems, the single-particle bands are nearly dispersionless, arising either from magnetic field-induced Landau levels or from band folding in moiré superlattices. The flatness of these bands amplifies the effects of electron-electron interactions, while their nontrivial topology enables the emergence of many-body states supporting fractionalized excitations. In the ideal flatband limit, the many-body Hamiltonian for partially filled bands takes the form \cite{lee2019theory,PhysRevResearch.3.L032070,PhysRevResearch.6.L032063}
\begin{equation}\label{eq:Hmb}
    H =\sum_\mathbf{q} V(\mathbf{q}) \hat{\rho}_\mathbf{q} \hat{\rho}_{-\mathbf{q}},
\end{equation}
where $\hat{\rho}_\mathbf{q}=\sum_{\mathbf{k}} \lambda_{\mathbf{q}}(\mathbf{k}) c^\dagger_{\mathbf{k}} c_{\mathbf{k+q}}$ is the density operator projected onto the flatband, and $\lambda_{\mathbf{q}}(\mathbf{k})\equiv \langle u_\mathbf{k}|u_{\mathbf{k+q}}\rangle$—the so-called form factor—encodes all geometric information of the Bloch wavefunction $|u_\mathbf{k}\rangle$. Unlike conventional correlated systems, where the Fermi surface plays a central role, here the focus is on interaction-driven physics governed by the structure of the single-particle wavefunctions themselves.

Traditionally, at fractional filling factors such as $\nu=1/3$, the lowest Landau level supports fractionally quantized Hall conductance $\sigma_{xy}=\nu e^2/h$ and quasiparticles with $\nu e$ fractional charge, enabled by the topological band with Chern number $|C|=1$. Recent theoretical and experimental advances have extended this paradigm to lattice flatbands with $|C|=1$, provided the Bloch wavefunctions exhibit sufficient analyticity. Analytical approaches to FQHS and FCIs leverage this analyticity—manifested as ideal quantum geometry \cite{PhysRevLett.127.246403,PhysRevResearch.5.023167,1zg9-qbd6,PhysRevResearch.2.023237,PhysRevLett.134.106502,fonseca2025gradient}, momentum-space holomorphicity \cite{PhysRevB.104.045103,PhysRevB.104.045104,PhysRevB.104.115160}, and related properties\cite{PhysRevB.108.205144}—to establish a direct correspondence with Landau level physics. These methods, rooted in the Girvin-MacDonald-Platzman (GMP) algebra \cite{PhysRevB.33.2481} and the commutator structure of projected density operators, enable powerful predictions for many-body states in the thermodynamic limit.

However, real materials and numerical studies rarely realize such ideal conditions. In practice, the quantum geometry is generally non-ideal, and analytical Bloch wavefunctions are unavailable. Moreover, in finite-size calculations such as exact diagonalization (ED), the many-body Hamiltonian does not access the Bloch wavefunctions in the whole Brillouin zone or their local momentum-space curvatures (quantum geometric tensor). Instead, the calculation relies on a discrete mesh of $\mathbf{k}$-points and the global correlations encoded in the form factors $\lambda_{\mathbf{q}}(\mathbf{k})$. Remarkably, even with these discretizations and truncations—finite system size, finite $\mathbf{k}$-mesh in the Brillouin zone, and limited momentum transfer $\mathbf{q}$—the essential features of fractionalization, such as ground state degeneracy and particle entanglement spectrum (PES) gaps, remain robust. This highlights the central role of form factors as the effective bridge between microscopic band structure and emergent many-body phenomena, both in analytical theory and in numerical simulations.

To move beyond analytical approaches and address the complexities of realistic, non-ideal quantum geometry, we adopt a numerical perspective focused on the form factors $\lambda_{\mathbf{q}}(\mathbf{k})$ obtained from finite-size single-particle flatbands (see Methods~\ref{med:formfactors} and SI \footnote{SI Sec. I: The original single-particle Hamiltonian and training form factors}). 
The form factor $\lambda_{\mathbf{q}}(\mathbf{k})$, which has much lower dimensionality than the full Bloch wavefunctions, serves as a compact descriptor of the quantum geometry in FCIs—much like the genome encodes essential biological information. Just as AlphaFold has transformed genomics by uncovering patterns and enabling predictions from complex genetic data \cite{jumper2021highly}, we anticipate that deep learning can similarly reveal hidden structures and generate new possibilities within the space of form factors, thereby advancing our understanding of correlated topological phases.

Given the central role of form factors as concise descriptors of quantum geometry, it is natural to ask how effectively machine learning can distinguish different many-body phases based on these objects. To address this, we first explore supervised classification approaches as an initial benchmark before turning to more advanced unsupervised methods. In Table S1, 
we apply supervised learning methods (logistic regression, fully connected neural networks, support vector machines \cite{bishop2006pattern,hastie2009elements}) to classify form factors corresponding to $\nu=1/3$ FCI states. Notably, the support vector machines \cite{cortes1995support} with linear kernels achieve high classification accuracy in the complex-valued feature space, indicating that the form factors are linearly separable in the original high-dimensional space. These results demonstrate that machine learning can effectively distinguish different quantum phases based on form factors.

Building on this foundation, we turn to unsupervised learning methods to uncover deeper physical insights and generative capabilities. In this work, we focus on two complementary approaches: variational autoencoders (VAE) \cite{kingma2013auto,goodfellow2016deep} and principal component analysis (PCA) \cite{bishop2006pattern,pearson1901liii,hotelling1933analysis}. The VAE, inspired by computer vision, treats form factors as two-dimensional images and learns to encode them into a low-dimensional latent space, from which new form factors can be generated by sampling. This generative modeling enables interpolation and extrapolation between different quantum geometries. To further interpret the structure of the learned latent space, we employ PCA to identify the principal directions of variance in the form factor dataset. PCA provides a linear decomposition that reveals the dominant modes underlying the data, offering a complementary perspective on the generalization power of the VAE and the emergence of form factors with different Chern numbers. In this sense, PCA acts as a renormalization group on the dataset variance when truncating form factors based on principal components, highlighting the most relevant features from the data.

\section{Results}

\subsection{Many-body Setting}

\begin{figure}[t!]
\begin{center}
\includegraphics[width = 0.48\textwidth]{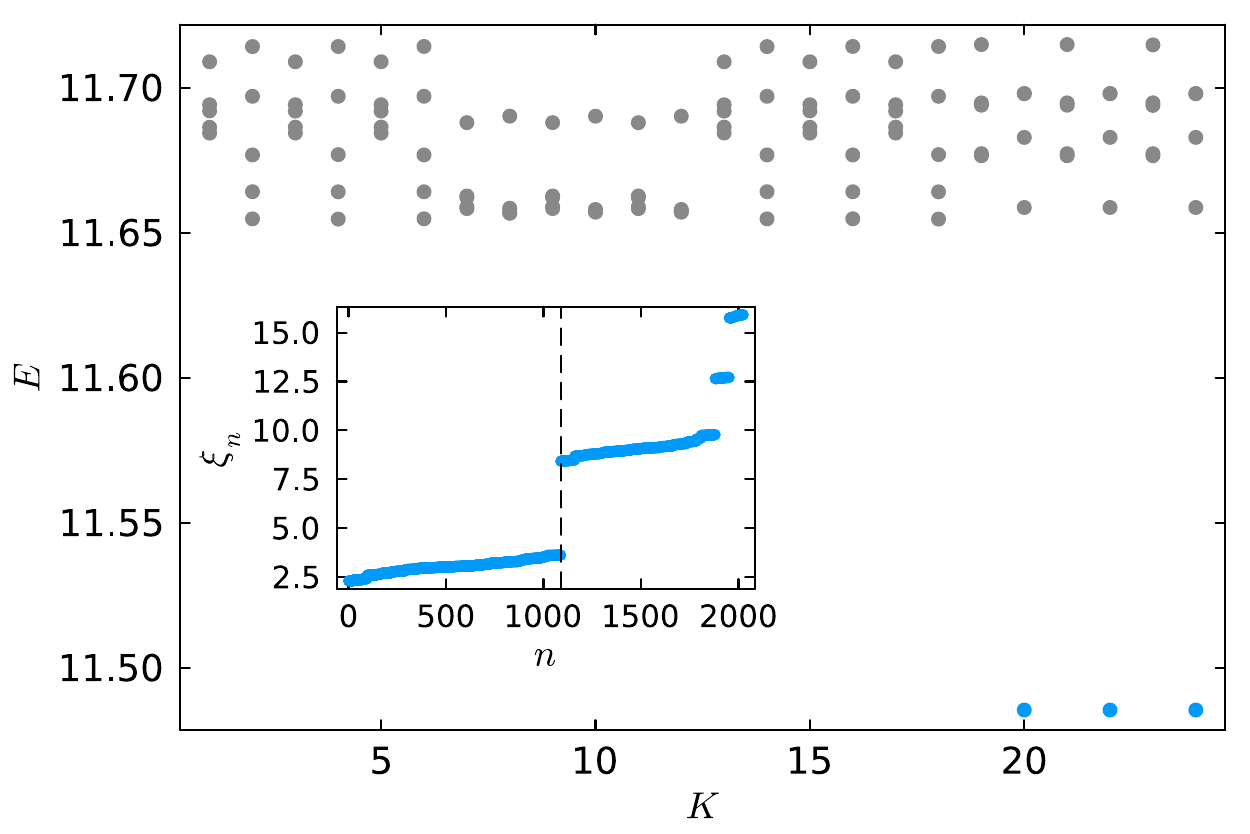}
\caption{\label{fig:fci} \textbf{Exact-diagonalization spectrum and particle entanglement spectrum (PES) of an ideal fractional Chern insulator (FCI).}
The ideal FCI many-body energy spectrum at the second magic parameter $\tilde{\alpha}_2=2.1325$ (see SI Sec. I) in the chiral limit with $c_0=0$, $N_x=4$, $N_y=6$ obtained at filling $\nu=1/3$. The ground states are labeled by their momentum sectors $K\equiv k_xN_y+k_y$. The FCI state is manifested as three quasi-degenerate ground states separated from other states with a many-body gap. It further exhibits a gap in the PES as shown in the inset. 
}
\end{center}
\end{figure}

To obtain the many-body ground and low-lying excited states, we solve the Hamiltonian Eq.~\ref{eq:Hmb} via ED calculation. The particle entanglement spectrum (PES) is also evaluated to diagnose the topological character of the ground states (\cite{PhysRevResearch.6.L032063} for details).

The many-body Hamiltonian in Eq.~\ref{eq:Hmb} is implemented on a finite momentum mesh of size $N_x \times N_y$ within the moiré Brillouin zone (BZ) \cite{PhysRevResearch.6.L032063}. Throughout this work, we focus on the case of $\nu=1/3$ filling for interacting spinless fermions. The form factors $\lambda_{\mathbf{q}}(\mathbf{k})$ are computed from a single-particle topological flatband model with Chern number $C=-1$, the quadratic band crossing point (QBCP) model—a low-energy two-band system with quadratic dispersion and a moir\'e periodic potential~\cite{PhysRevLett.130.216401} ($H_0(\tilde{\alpha},c_0)$ see Methods~\ref{med:formfactors} and SI Sec. I). These form factors are complex-valued and defined on discrete meshes of both $\mathbf{k}$ and $\mathbf{q}$, with the momentum transfer $\mathbf{q}$ truncated to within $127$ BZs for computational convergence of the many-body calculation. The choice of interaction $V(\mathbf{q})$ plays a crucial role in stabilizing many-body phases. In this study, we primarily use a screened Coulomb interaction $V(\mathbf{q}) = 4\pi \tanh(\sqrt{3}q/(4\pi))/(\sqrt{3}q)$ \cite{PhysRevResearch.3.L032070}, which decays slowly in $q$ and yields robust FCI states in both ideal and non-ideal form factors \footnote{While a contact interaction $V(\mathbf{q}) = -q^2$ produces exact zero-energy FCI ground states for $\nu=1/3$ in the ideal case \cite{PhysRevResearch.5.023167}, it is unbounded from below and leads to instability for non-ideal form factors. In contrast, interactions with fast-decaying in $q$, i.e., with $\tanh(dq), d \gg 1$ tend to destabilize the FCI phase}.

An example of the many-body energy spectrum and PES for an ideal FCI state is shown in Fig.~\ref{fig:fci}. The ground state manifold exhibits threefold quasi-degeneracy in specific momentum sectors, and the PES shows a gap at the $(1,3)$-admissible counting for partition $Q_A$ (number of particles in subsystem A), consistent with the expected signatures of a $\nu=1/3$ FCI state. To label the ground states as FCI, we adopt the following criteria: (1) the ground state manifold exhibits threefold quasi-degeneracy in specific momentum sectors at $\nu=1/3$ filling; (2) the PES shows a gap at the $(1,3)$-admissible counting; and (3) the many-body gap $\Delta = E_4 - E_3$ exceeds the ground state splitting $D = E_3 - E_1$. We assign an FCI label of $1$ if all criteria are satisfied, and $0$ otherwise. 
The third criterion is a conservative finite-size requirement that selects samples with a well-separated quasi-degenerate ground-state manifold, thereby reducing ambiguous labels caused by large finite-size splitting.

\subsection{Variational Autoencoder (VAE)}

\begin{figure*}[t]
\begin{center}
\setlabel{pos=nw,fontsize=\large,labelbox=false}
\xincludegraphics[scale=0.5,label=a]{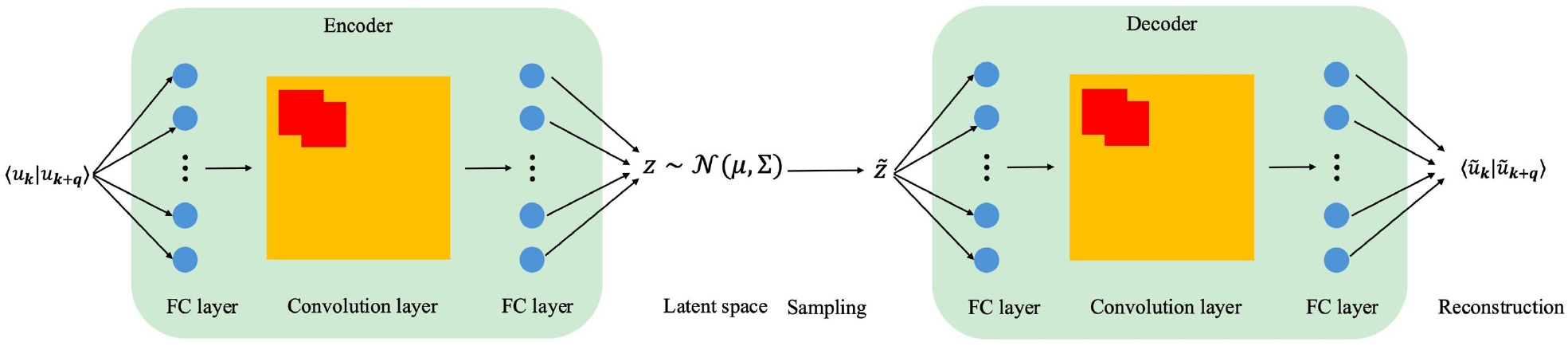}
\xincludegraphics[scale=0.3,label=b]{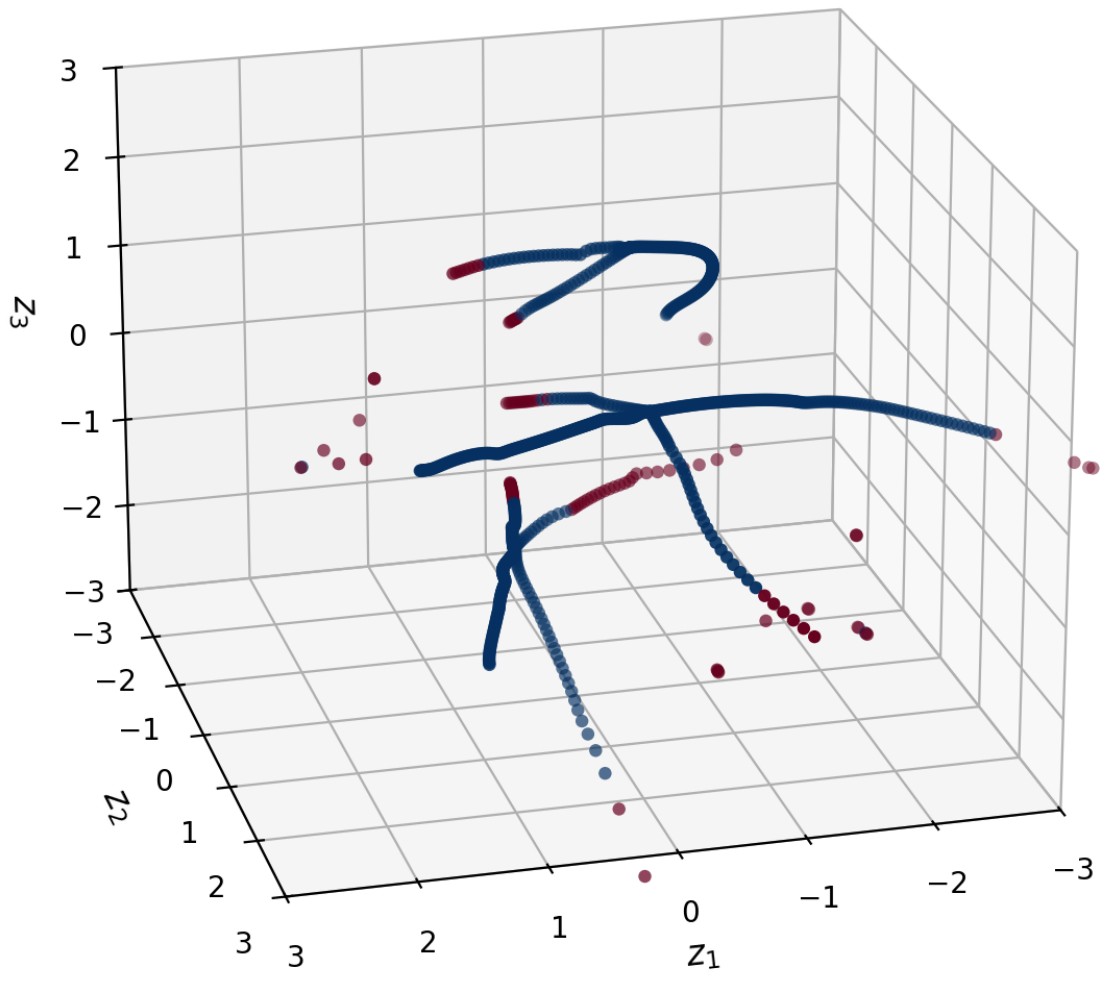}
\xincludegraphics[scale=0.35,label=c]{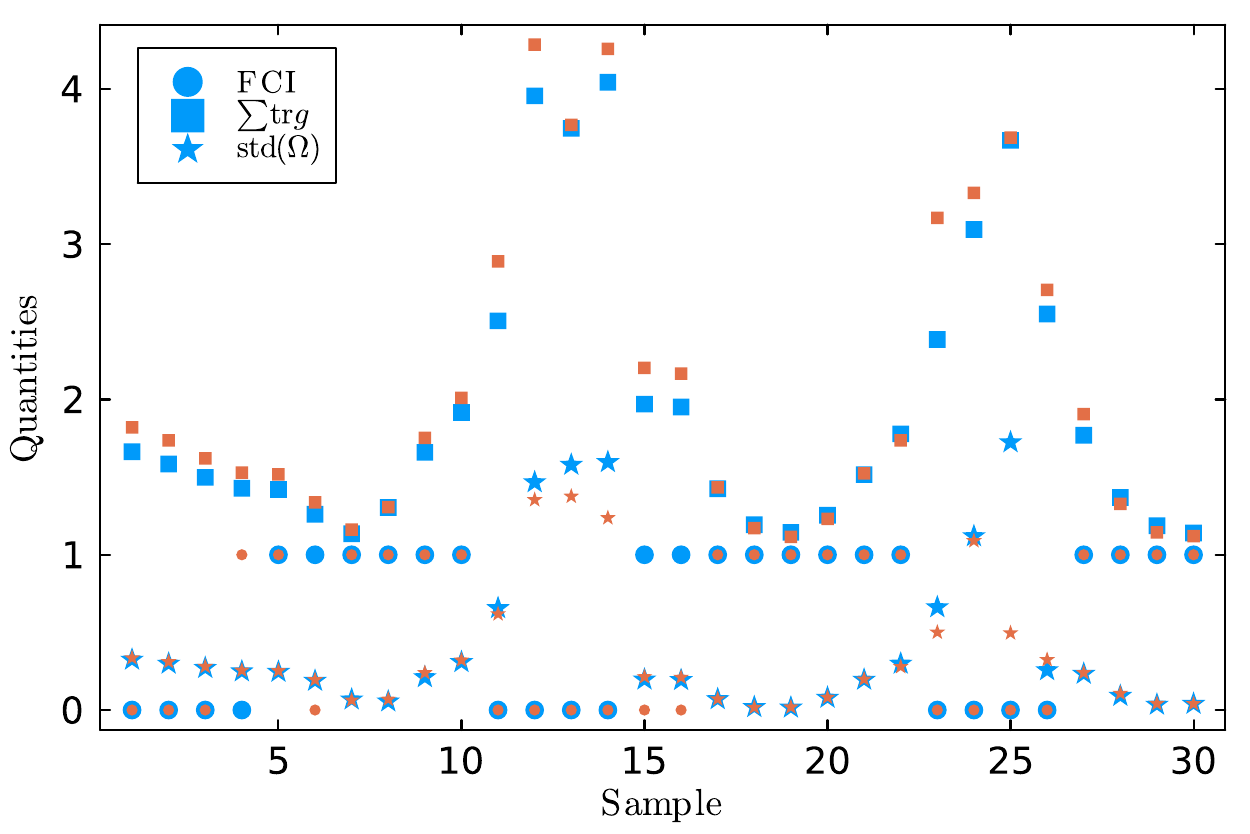}
\caption{\label{fig:demo} \textbf{Variational autoencoder (VAE) architecture, latent-space organization, and reconstruction performance for form factors.}
(a) Schematic of the VAE architecture. (b) Three-dimensional latent space positions of all training data at their mean values. System size: $N_x=4$, $N_y=6$. The model is trained with all form factors. Blue dots indicate fractional Chern insulator (FCI) samples, while red dots indicate non-FCI samples. (c) Comparison between original and generated form factors from the corresponding latent space mean values $z$ in terms of total quantum metric $\sum \mathrm{tr} g$, Berry curvature deviation $\mathrm{std}(\Omega)$, and the many-body ground state at $\nu=1/3$ (1 for FCI states and 0 for non-FCI states). Blue symbols represent the original form factors, and red symbols represent the generated ones.
}
\end{center}
\end{figure*}

The variational autoencoder (VAE)~\cite{kingma2013auto} 
has recently found increasing applications in physics~\cite{samarakoon2020machine,rocchetto2018learning,yin2021neural,PhysRevB.105.224205,PhysRevE.96.022140,hou2024unsupervised,kalinin2022atomically}. The core idea is to learn a probabilistic mapping between high-dimensional data and a lower-dimensional latent space $z = (z_1, \cdots, z_L)^T$, where, across the training data, $z$ is assumed to follow a multivariate Gaussian distribution $Z \sim \mathcal{N}(\mathbf{0},\mathcal{I})$. In physics terms, this is analogous to identifying a minimal set of collective variables or order parameters that capture the essential physics of a complex system. The VAE learns the underlying distribution of the training data, with marginal likelihood is given by
\begin{equation}
    P_\theta(x) = \int P_\theta(x|z) P_\theta(z) \mathrm{d}z,
\end{equation}
where $\theta$ denotes the parameters of the neural networks.

A VAE consists of two main components: an encoder, which maps each input to a point (with uncertainty, posterior density $P_\theta(z|x)=P_\theta(x|z)P(z)/P_\theta(x) \sim \mathcal{N}(\mu(x),\Sigma(x))$) in the latent space, and a decoder, which reconstructs the original data from this latent representation. While both the marginal likelihood $P_\theta(x)$ and the posterior density $P_\theta(z|x)$ are intractable~\cite{kingma2013auto}, 
the probabilistic structure of the VAE ensures that the latent variables are uncorrelated and that the model can generate new, physically plausible form factors by sampling the latent space. 
The training objective balances reconstruction accuracy, the Kullback-Leibler (KL) divergence~\cite{kullback1951information} in the latent space, and model parameter regularization, ensuring that the learned representation captures the essential features of the data while maintaining generative capability.

Fig. ~\ref{fig:demo}a illustrates our VAE architecture, which uses fully connected (FC) layers at the input/output and latent space, with a convolutional layer \cite{convolution} in the middle for both encoder and decoder. The form factor is reshaped into a 2D matrix for the convolution, which empirically yields the best performance. We also tested deeper FC networks and transformer layers (see Table S2), 
but found that the convolutional architecture is most effective for capturing the local correlations in hidden space of form factors. 

When encoding the training data, the encoder maps all samples into a three-dimensional latent space, forming roughly three connected curves (corresponding to the first three magic parameters of the QBCP model), while non-FCI points are more scattered, as shown in Fig. ~\ref{fig:demo}b. Notably, the FCI labels were obtained from ED and not provided during training.  Therefore, the clear separation of FCI and non-FCI samples in latent space demonstrates the effectiveness of the learned representation (Figure S4a). The reconstructed form factors closely match the originals (The average reconstruction losses are shown in SI Sec. IV.) in terms of total quantum metric $\sum \mathrm{tr} g$, Berry curvature deviation $\mathrm{std}(\Omega)$, and the many-body ground state at $\nu=1/3$ as shown in Fig.~\ref{fig:demo}c. 

\begin{figure}[t!]
\begin{center}
\setlabel{pos=nw,fontsize=\large,labelbox=false}
\xincludegraphics[scale=0.35,label=a]{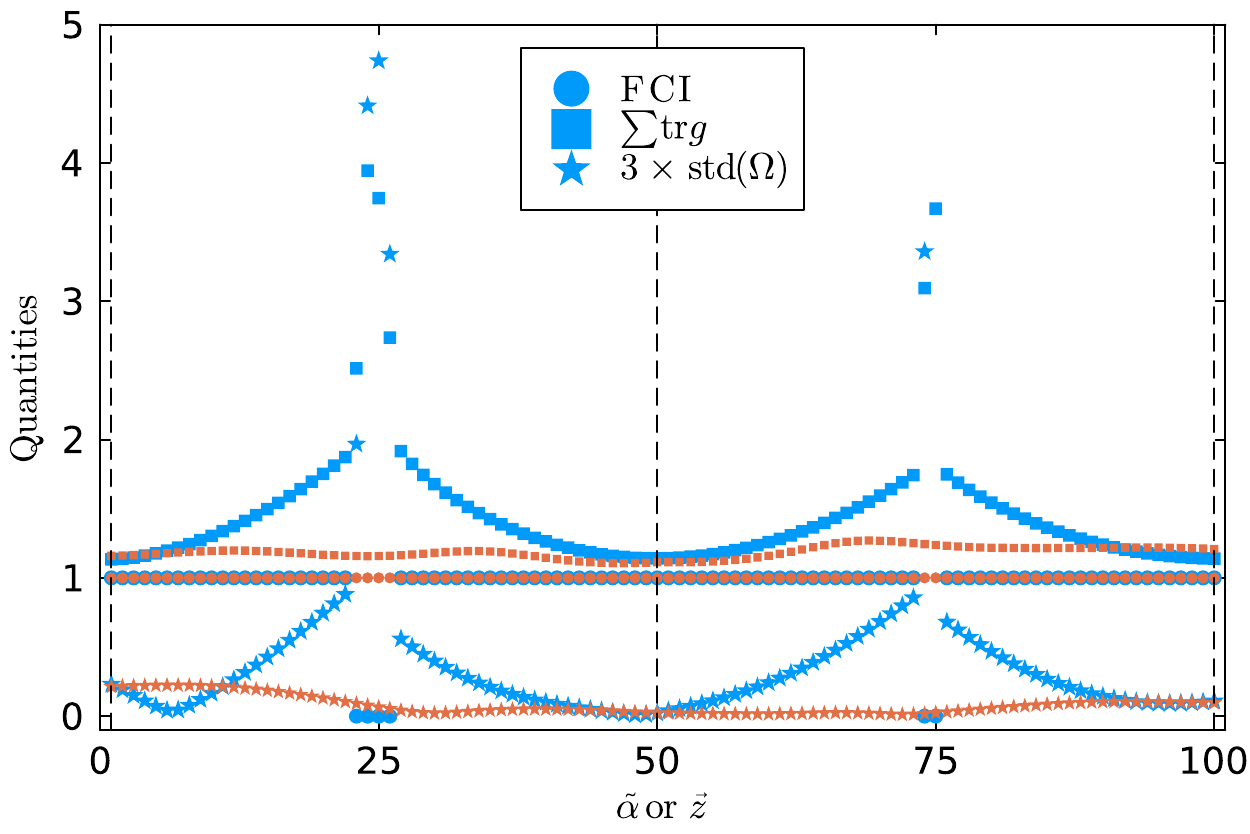}
\xincludegraphics[scale=0.35,label=b]{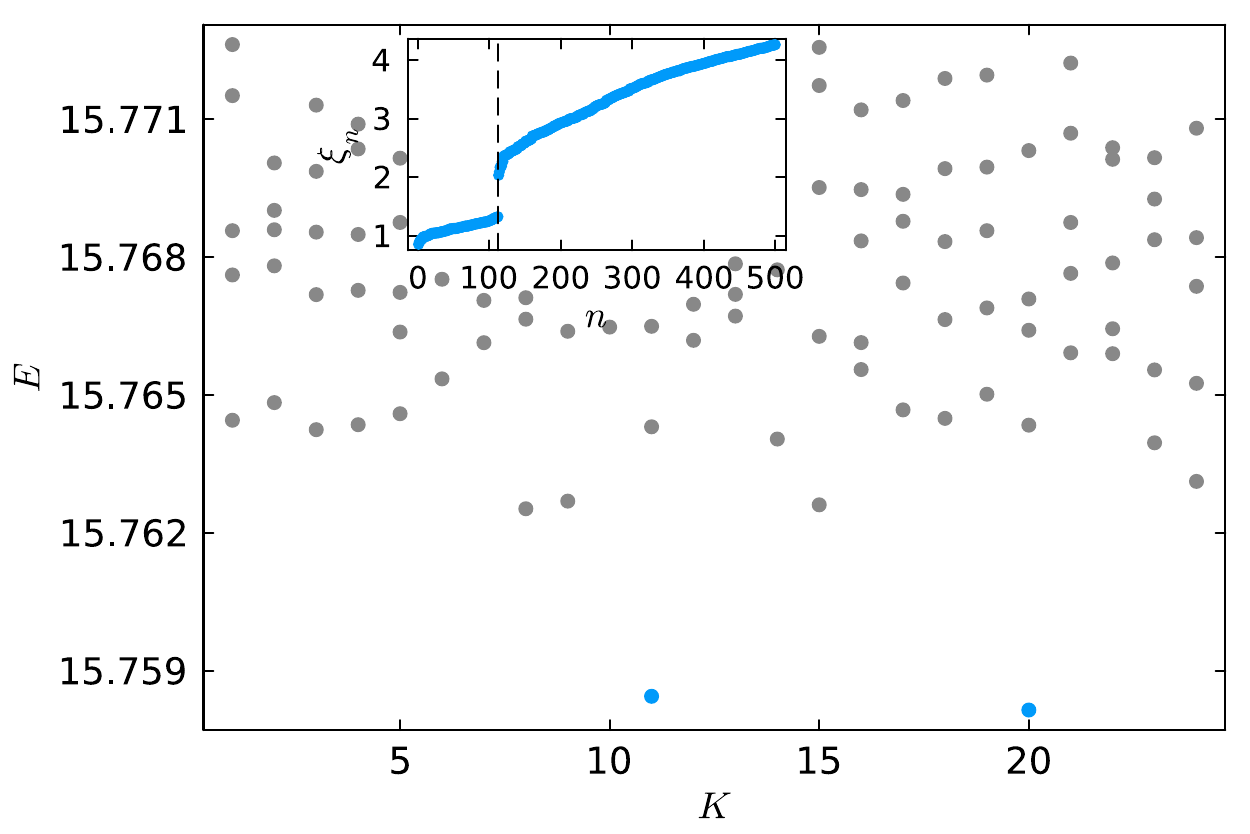}
\xincludegraphics[scale=0.35,label=c]{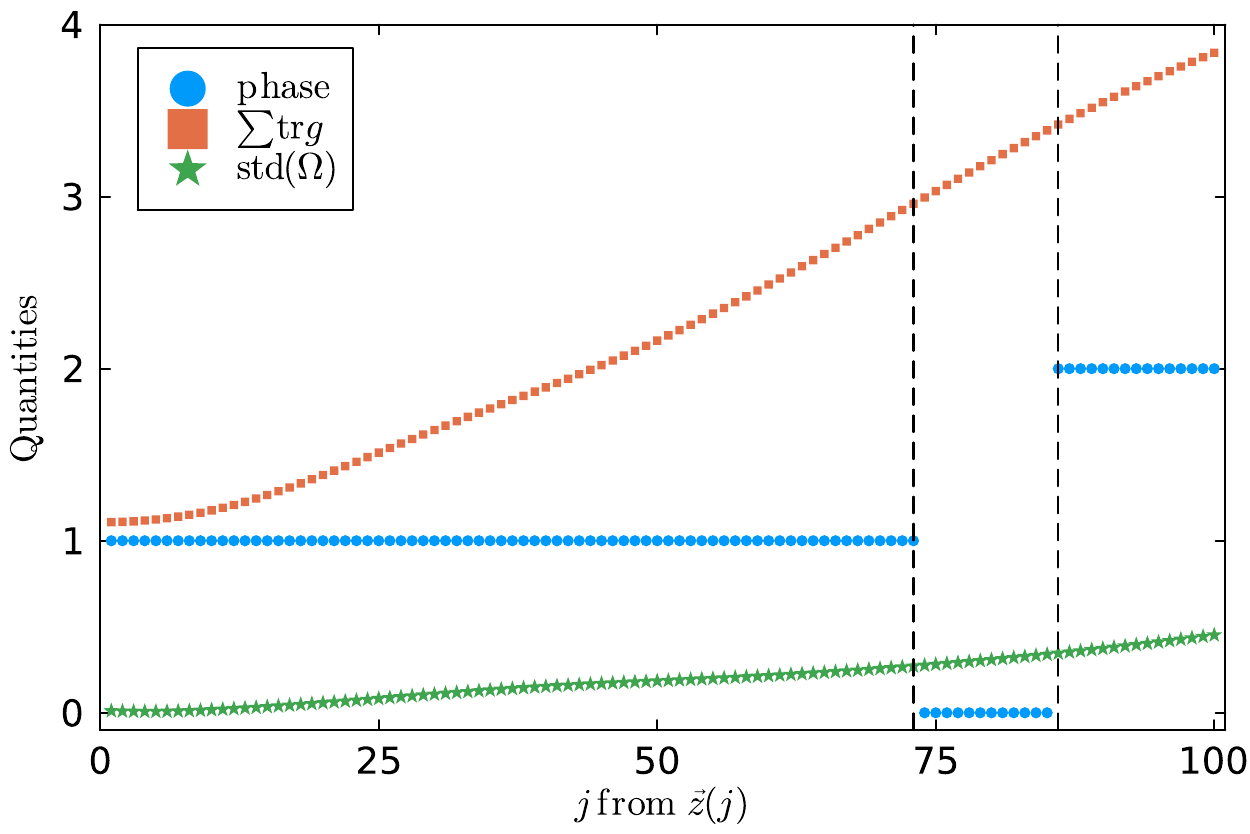}
\caption{\label{fig:extrapolate}
\textbf{Latent-space interpolation and extrapolation generate fractional Chern insulator (FCI) and charge-density-wave (CDW) states.}
Modeling quantum many-body states in the latent space using the variational autoencoder for system size $N_x \times N_y = 4 \times 6$. (a) The FCI label, total quantum metric, and Berry curvature variance from the original single-particle model as a function of $\tilde{\alpha}$ (blue symbols), compared to those obtained by interpolating between three ideal cases (magic parameters) in the latent space (red symbols). (b) A generated form factor sample from the latent space that yields two degenerate CDW ground states. Inset: the particle entanglement spectrum shows a gap for the charge density wave, where the number of quasihole excitations is the number of ways of placing $Q_A=3$ particles into $Q=8$ sites times the number of charge ordered states, $2\binom{Q}{Q_A} = 112$ (see Supplementary Information in Ref. \cite{43nq-ntqm}). (c) Many-body results when interpolating between FCI (at the second magic parameter) and CDW samples in the latent space. Phase: $1$ for FCI states, $2$ for CDW states, $0$ for crossover states.
}
\end{center}
\end{figure}

A key advantage of the latent space is that it enables smooth interpolation between different ``ideal'' form factors at various magic parameters, without closing the single-particle gap as would occur in the original single-particle model(Figure S4b). As shown in Fig.~\ref{fig:extrapolate}a, interpolating between form factors at the three magic parameters in latent space preserves the quantized quantum metric and robustly yields FCI ground states in ED. In contrast, varying the single-particle parameter $\tilde{\alpha}$ in the original model leads to gap closures and large variations in quantum geometry. Thus, the VAE latent space provides a flexible model space for mixing features of different quantum geometries, potentially enabling new ways to tune form factors between analytical models such as Landau levels and chiral moiré bands.

The generative power of the VAE also allows extrapolation: by sampling latent vectors $z$ beyond the training range, we can generate new form factors with quantized Chern number and varying quantum metric. By filtering generated samples for $C=-1$ and high quantum metric, we identify form factors that yield charge density wave (CDW) ground states, which are not accessible in the original single-particle model [Fig.~\ref{fig:extrapolate}b and Figure S4c]. 
This feature points to future work on models closer to the first Landau level or generalized Landau levels \cite{1zg9-qbd6} to target non-Abelian phases, which are promising for fault-tolerant quantum computing applications and deepen our understanding of strongly correlated topological matter.

Furthermore, the latent space enables continuous interpolation between FCI and CDW form factors [Fig.~\ref{fig:extrapolate}c], revealing a phase evolution region without leaving the $C=-1$ manifold \footnote{The spectra of the evolution are shown in SI. Sec. IV Fig. 4}. This provides a new way to model and explore phase boundaries between different correlated states, and could be extended to connect Abelian and non-Abelian FCIs.

Interestingly, even when training only on $C=-1$ FCI form factors, 
the VAE can generate $C=0$ form factors, indicating that the model can generalize beyond the training topological class. This highlights the unique perspective offered by machine learning compared to traditional approaches based on Bloch wavefunctions and local curvature. In the next section, we use principal component analysis (PCA) to further investigate the global correlations and structure of the form factor data.

Finally, one may ask whether it is possible to invert the VAE and recover the physical parameters of the original single-particle model from a given form factor. In principle, this could be approached by modifying the loss function to include a term penalizing deviation from known physical parameters. For form factors close to the training set, this may be feasible, but the mapping from physical parameters to form factors is highly nonlinear and generally not one-to-one, especially for form factors not realizable in the original model. 
When the generated form factors cannot be realized by the original single-particle model, their physical interpretation may be unclear. However, the fact that these form factors can yield FCI or CDW states in ED suggests they are physically meaningful within the context of the many-body Hamiltonian in Eq.~\ref{eq:Hmb}. Their associated behaviors in Berry curvature and quantum metric provide valuable insights into the relationship between quantum geometry and many-body physics, offering a broader perspective beyond the original single-particle model. The ability of the VAE to learn and generate physically relevant form factors paves the way for future work on the inverse design of form factors with tailored quantum geometry and many-body phases.


In practice, the architecture of the VAE also plays a crucial role. Different neural network architectures yield substantially different performance (SI Sec. IV), especially for smaller network sizes and limited computational resources. In our tests, the hybrid architecture combining fully connected layers with a convolutional layer was the most accurate and stable. This likely reflects the structured momentum-space correlations of the form factors: the fully connected layers provide a flexible global representation, while the convolutional layer is effective at extracting local patterns after reshaping the data. More generally, these results indicate that physically informed architectural choices can significantly improve learning for topological quantum-geometry data.

\subsection{Principal Component Analysis (PCA) of Form Factors}

\begin{figure}[t]
\begin{center}
\setlabel{pos=nw,fontsize=\large,labelbox=false}
\xincludegraphics[scale=0.36,label=a]{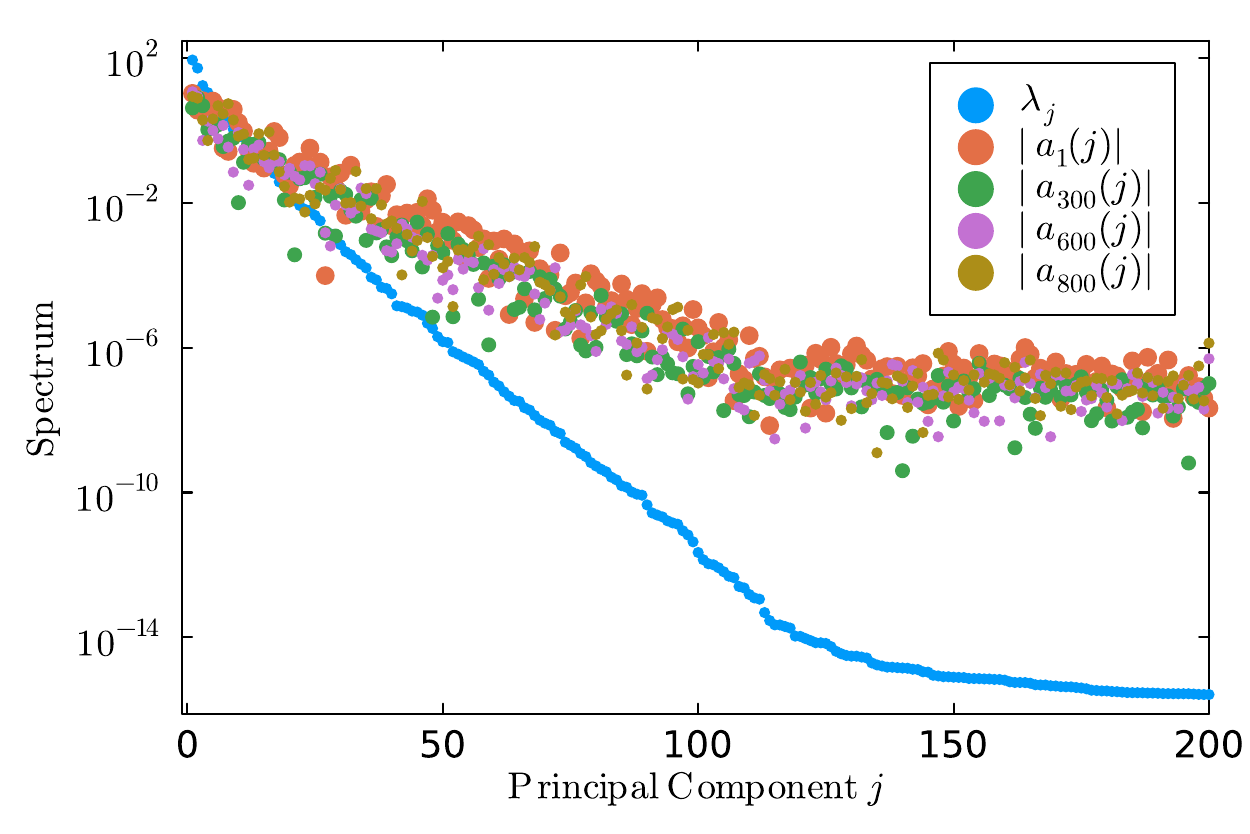}
\xincludegraphics[scale=0.35,label=b]{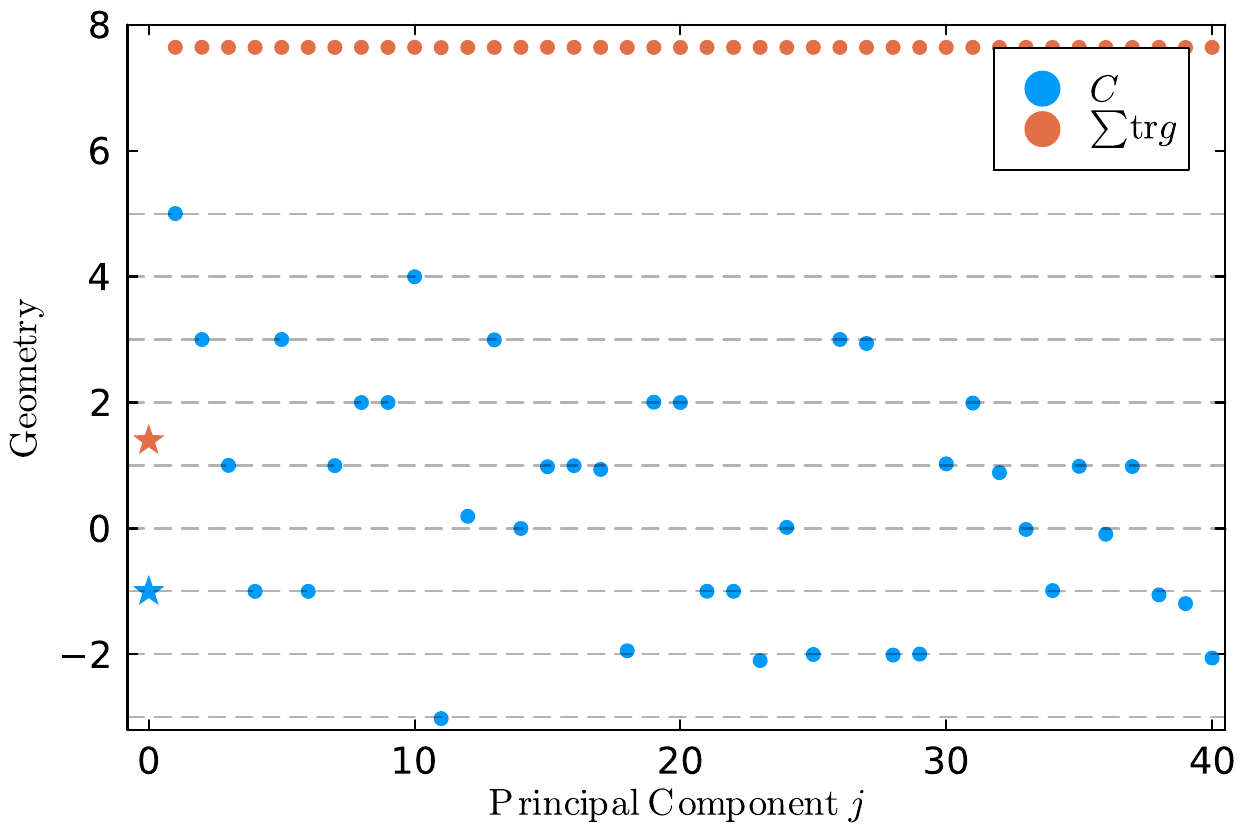}
\caption{\label{fig:pca} 
\textbf{Principal component analysis (PCA) identifies dominant variance modes and their associated quantum geometry.}
PCA analysis of the form factors for system size $N_x \times N_y = 4 \times 6$, using only fractional Chern insulator (FCI) samples. (a) Eigenvalues (variance) of the covariance matrix and the absolute coefficients $a_n(j)$ from the PCA expansion for four example samples, where $n$ is the training sample index. (b) Quantum geometry (Wilson loop Chern number and total quantum metric) of the first $40$ principal components. Star symbols at index 0 indicate the corresponding values for the mean of the training data.
}
\end{center}
\end{figure}

To further understand the emergence of $C=0$ form factors generated by the VAE, we apply principal component analysis (PCA) as an unsupervised learning tool to the form factor dataset. PCA extracts the principal directions of variance in the data, providing an efficient linear basis for representing the complex-valued form factors. After determining the principal components, each data point $\mathbf{x}_n$ can be expanded in terms of a truncated expansion in the first $M$ components $\mathbf{u}_j$ around the mean $\bar{\mathbf{x}}$:
\begin{equation}
    \tilde{\mathbf{x}}_n = \bar{\mathbf{x}} + \sum_{j=1}^M a_{nj} \mathbf{u}_j,
\end{equation}
where $a_{nj}$ are the expansion coefficients (with $n$ indexing the data samples and $j$ the principal components; see Methods~\ref{med:PCA} for details).

As shown in Fig.~\ref{fig:pca}a, the eigenvalues of the covariance matrix exhibit exponential decay, indicating that only a small number of principal components account for the majority of the variance in the data. This rapid decay primarily arises from the linear structure when the form factors are represented by their real and imaginary parts \footnote{When using phase and amplitude instead (or taking a log of the form factors), the decay is less pronounced due to increased nonlinearity in the data representation.}. Correspondingly, the expansion coefficients for each training form factor in the principal component basis also decrease quickly, enabling efficient truncation. In practice, retaining approximately $20$ principal components suffices to accurately reconstruct the original form factors, as verified by ED results (SI Sec.~V). For non-FCI form factors, projecting onto the truncated principal component basis centered on FCI samples can even enhance FCI-like characteristics \footnote{SI Sec.~V: PCA expansion and ED results, Figure S6.}.

More intriguingly, examining the quantum geometry of individual principal components [Fig.~\ref{fig:pca}b] reveals that their Chern numbers, computed via Wilson loop calculations at the central Brillouin zone, are approximately quantized values ranging from $-3$ to $5$ for the $C=-1$ training data. Note that the principal components are not universal objects, but data-dependent directions determined by the covariance structure of the training set. They are therefore not intrinsic physical states and are not guaranteed to have quantized Chern numbers, partly because 
$\langle u_\mathbf{k}|u_{\mathbf{k+G}} \rangle \neq 1$. Nevertheless, the approximately quantized Chern numbers observed for many leading components suggest that deviations from ideal quantum geometry can be interpreted as superpositions of variance modes with different topological character.
Importantly, the Chern number of a reconstructed form factor is not simply the sum of the Chern numbers of its principal components, reflecting the highly non-linear and nonlocal nature of the form factor as an object derived from Bloch wavefunctions. The small amplitude variations of form factors $\delta\lambda$ are captured by large total quantum metric of the principal components in Fig. \ref{fig:pca}b, where the amplitude of each component is close to $0$.
These findings also highlight the nontrivial structure underlying the VAE results: when varying form factors with $\tilde{\alpha}$, global topological variations must be considered. Similar trends are also observed when applying PCA to the full dataset (including non-FCI samples). Although the physical meaning of principal components with higher Chern numbers remains unclear, they nonetheless provide an accurate expansion of the original form factors for both FCI and non-FCI samples (see Figures S6 and S7).

From a machine learning perspective, PCA provides a compact representation of the form factors, reducing each sample to a set of coefficients in the principal component basis. However, training neural networks or VAEs directly on these coefficients performs poorly, as the essential raw features are largely washed out by the linear transformation. 
Gaussian mixture models can be used to sample the coefficient space and reconstruct new form factors; we have found that this approach can generate physically plausible form factors. However, the high dimensionality and lack of interpretability of the principal component space present significant challenges, which we leave for future work.

\section{Discussion}

In summary, we have demonstrated that unsupervised machine learning methods, specifically VAE and PCA, provide a powerful framework for modeling and generating form factors in fractional Chern insulators. By learning directly from the distribution of form factors, our approach enables both interpolation between ideal quantum geometries and extrapolation to new many-body states, including charge density waves, without explicit reliance on analytical wavefunctions. The PCA analysis further reveals that global correlations in form factors can be decomposed into components with approximately quantized Chern numbers, offering new insights into the structure of quantum geometry in topological flatbands.

Importantly, our results show that even form factors not realizable in the original single-particle model can yield physically meaningful many-body states, as evidenced by the emergence of FCI or CDW phases in ED calculation. This highlights the broader perspective afforded by machine learning, which can generalize beyond traditional approaches based on Bloch wavefunctions and local curvature, and paves the way for the inverse design of form factors with tailored quantum geometry and many-body phases.

Our results open several promising directions for future research. First, while this study focuses on Abelian FCI phases, it would be highly valuable to extend our framework to non-Abelian FCIs—such as the Moore-Read (MR)~\cite{MOORE1991362} and Read-Rezayi (RR)~\cite{PhysRevB.59.8084} states—by training on models that more closely resemble the first Landau level or its generalizations. Second, our findings underscore the crucial role of neural network architecture in capturing the physical symmetries and correlations inherent to topological systems. Future work could systematically explore hybrid or symmetry-adapted architectures, for example by combining convolutional and attention-based layers, to further improve the learning of topological features. Third, our observation that randomly generated form factors—far from the ideal limit—can still stabilize FCI phases points to a new avenue for understanding the robustness of fractionalization. The machine learning framework developed here could be leveraged to systematically explore the space of form factors that promote the emergence of FCIs.

\section{Methods}

\subsection{Form factors as training data and discrete geometry}
\label{med:formfactors}

The choice and quality of training data are crucial for effective machine learning applications in physics. Unlike many statistical problems, physical training data are typically less affected by random noise and are grounded in well-defined models. In this work, we generate training data from a physical single-particle Hamiltonian—the quadratic band crossing point (QBCP) model~\cite{PhysRevLett.130.216401}—which provides both analytically tractable (ideal) and non-ideal form factors. Details of the QBCP model can be found in Ref.~\cite{PhysRevLett.130.216401}. Briefly, the training form factors are derived from the single-particle Hamiltonian $H_0(\tilde{\alpha},c_0)$, where $\tilde{\alpha}$ is a dimensionless parameter controlling the moir\'e potential strength, and $c_0$ tunes the quantum metric anisotropy of the two central flatbands via the low-energy quadratic dispersion. The training set includes form factors from the lower flatbands of $H_0(\tilde{\alpha},0)$ and $H(\tilde{\alpha}_j,c_0)$, where $\tilde{\alpha}_j$ ($j=1,2,3$) are the first three magic parameters (SI Sec. I). At these magic parameters, the QBCP yields analytically exact flatbands with ideal quantum geometry. For $c_0=0$, the model exhibits sublattice chiral symmetry. The resulting dataset contains both ideal and non-ideal bands, predominantly with Chern number $C=-1$, but also includes some with $C\neq -1$ as $\tilde{\alpha}$ is varied.

Each form factor $\lambda_\mathbf{q}(\mathbf{k})$ is stored as a two-dimensional array. In principle, both the momentum transfer $\mathbf{q}$ and the momentum point $\mathbf{k}$ can span multiple Brillouin zones (BZs). However, due to translational symmetry, $\lambda_{\mathbf{q}}(\mathbf{k+G}) = \lambda_{\mathbf{q+G}}(\mathbf{k})$, we can restrict $\mathbf{k}$ to the first BZ and allow $\mathbf{q}$ to extend over nearby BZs, subject to a truncation cutoff. Thus, each form factor is represented as an array of size $2N_\mathrm{BZ}N_k^2$, where $N_k$ is the number of momentum points in a single BZ, $N_\mathrm{BZ}$ is the number of BZs included, and the factor of $2$ accounts for the real and imaginary parts. Specifically, each form factor is stored as a $2N_k \times (N_k N_\mathrm{BZ})$ array, with the first $N_k$ rows for the real part and the remaining for the imaginary part. In this work, we use $N_k = 4 \times 6$ and $N_\mathrm{BZ} = 127$, resulting in an input dimension of $146{,}304$, which is much larger than the number of training samples.

The Bloch wavefunction possesses a gauge degree of freedom: diagonalizing the single-particle Hamiltonian yields $|u_\mathbf{k}\rangle \to e^{i\phi_\mathbf{k}}|u_\mathbf{k}\rangle$. This induces a corresponding gauge freedom in the form factors,
\begin{equation}
    \lambda_\mathbf{q}(\mathbf{k}) = \langle u_{\mathbf{k}}|u_{\mathbf{k+q}}\rangle \to e^{i(\phi_{\mathbf{k+q}}-\phi_{\mathbf{k}})}\langle u_{\mathbf{k}}|u_{\mathbf{k+q}}\rangle.
\end{equation}
Thus, form factors themselves are not gauge-invariant. To regularize this freedom, we fix the gauge by referencing each Bloch wavefunction to that at $\mathbf{k}=0$:
\begin{equation}
    |\bar{u}_\mathbf{k}\rangle \equiv \frac{\langle u_\mathbf{k}| u_0\rangle}{|\langle u_\mathbf{k}| u_0\rangle|} |u_\mathbf{k}\rangle.
\end{equation}
Note that if the training data are generated in a particular gauge, a randomly gauged form factor will appear as an outlier in the latent space. Consequently, the VAE cannot learn gauge-independent features of the interacting Hamiltonian, as the training data themselves are gauge-dependent. Visualization of form factors can be found in SI Sec. I. Developing deep learning methods that can extract gauge-invariant features remains an interesting direction for future work.

With the gauge fixed, we characterize form factors by connecting the discrete data to continuum quantum geometry. The local curvature is described by the quantum geometric tensor,
\begin{equation}
    \eta_{\mu\nu}(\mathbf{k}) \equiv \langle \partial_\mu u_\mathbf{k}|[1-|u_\mathbf{k}\rangle \langle u_\mathbf{k}|] |\partial_\nu u_\mathbf{k}\rangle,
\end{equation}
whose imaginary part gives the Berry curvature, $\Omega(\mathbf{k}) = -2\Im [\eta_{xy}(\mathbf{k})]$, and whose real part yields the Fubini-Study metric, $g_{\mu\nu}(\mathbf{k}) = \Re[\eta_{\mu\nu}(\mathbf{k})]$. The Berry curvature provides a lower bound for the quantum metric, $\mathrm{tr}\,g(\mathbf{k}) = g_{xx}(\mathbf{k}) + g_{yy}(\mathbf{k}) \geq |\Omega(\mathbf{k})|$, with the ideal quantum metric saturating this bound across the BZ.

Direct finite-difference calculations of these quantities are not gauge-invariant and can be numerically unstable. To ensure gauge invariance, we employ the Wilson loop approach to compute Berry phases and parallel transport. For the Berry curvature, $\Omega(\mathbf{k}) = i(\langle \partial_x u|\partial_y u\rangle - \langle \partial_y u|\partial_x u\rangle) = \nabla_\mathbf{k} \times \mathbf{A}$, where $\mathbf{A} = i\langle u|\nabla_\mathbf{k}|u\rangle$ is the Berry connection, the Wilson loop yields
\begin{equation}
\begin{split}
    C &= \frac{1}{2\pi} \int_{BZ} \Omega(\mathbf{k})\, d^2k = \frac{1}{2\pi} \oint_C \mathbf{A}(\mathbf{k}) \cdot d\mathbf{k} \\
    &= \frac{1}{2\pi} \sum_{j} \oint_{C_j} \mathbf{A}(\mathbf{k}) \cdot d\mathbf{k} \\
    &= \frac{1}{2\pi} \sum_{j} \sum_{\mathbf{k}_j \in C_j} i(\langle u_{\mathbf{k}_j}|u_{\mathbf{k}_{j+1}}\rangle - 1) = \frac{1}{2\pi} \sum_{j} -\phi_j,
\end{split}
\end{equation}
where $\langle u_{\mathbf{k}_j}|u_{\mathbf{k}_{j+1}}\rangle \approx e^{i\phi_j}$. Thus, the total Berry phase is obtained by summing the phases accumulated around each loop, and the Berry curvature can be approximated as
\begin{equation}
    \Omega(\mathbf{k}_j) S_j \approx -\phi_j,
\end{equation}
where $\mathbf{k}_j$ is the central momentum of loop $C_j$ and $S_j$ is the area enclosed by the loop. The sign of $\phi_j$ depends on the loop orientation. The total Chern number is then $C = \sum_j \Omega(\mathbf{k}_j) S_j / (2\pi)$, which is quantized.

Similarly, the quantum metric can be estimated from the squared overlap of Bloch states,
\begin{equation}
    s^2(\mathbf{k},\mathbf{k+q}) \approx 1 - |\langle u_\mathbf{k}|u_{\mathbf{k+q}}\rangle|^2 \underset{q\to 0}{\approx} g_{\mu\nu} q_\mu q_\nu,
\end{equation}
and evaluated in Wilson loops. For a simple square loop with steps $q_x, q_y$, the local trace of the metric is
\begin{equation}
    \mathrm{tr}[g(\mathbf{k}_j)] S_j = (g_{xx} q_x^2 + g_{yy} q_y^2) \approx \sum_{\mathbf{k}_j \in C_j} \frac{s^2(\mathbf{k}_j, \mathbf{k}_{j+1})}{2},
\end{equation}
where $\mathbf{q}_j = \mathbf{k}_{j+1} - \mathbf{k}_j$ and typically $q_x = q_y = q$. In the ideal limit, $\mathrm{tr}(g) = F_{xy}$, so this quantity should be approximately quantized to $2\pi$, up to finite-difference errors. Alternatively, the local metric can be estimated by evaluating quantum distances to nearest neighbors (see SI), but the Wilson loop approach is more closely related to the Berry curvature calculation. In the main text, we denote $\sum \mathrm{tr} g \equiv \sum_j \mathrm{tr}[g(\mathbf{k}_j)] S_j / (2\pi)$, which ideally approaches the Chern number $|C|$. The benchmarking of these discrete geometry calculations can be found in SI Sec. II.

\subsection{Variational Autoencoder}\label{med:VAE}

The variational autoencoder (VAE) is a generative model that learns a probabilistic mapping between high-dimensional data and a lower-dimensional latent space. The VAE consists of two neural networks: an encoder and a decoder. The encoder maps each input $X$ to a latent space distribution, parameterized by a mean $\mu$ and covariance $\Sigma$, i.e., $z(X) \sim \mathcal{N}(\mu, \Sigma)$. The decoder reconstructs the original data from a sample $z$ drawn from this distribution.

To regularize the latent space and enable generative sampling, a standard normal prior $z \sim \mathcal{N}(0, \mathcal{I})$ is imposed. The training objective is to maximize the likelihood of the data under the model, which is equivalent to minimizing the sum of two loss terms:
\begin{enumerate}
    \item \textbf{Reconstruction loss:}
    \begin{equation}
        L_\mathrm{recon}(x_n) = \sum_j (x_{nj} - \tilde{x}_j)^2,
    \end{equation}
    where $x_n$ is the $n$th training sample, $x_{nj}$ its $j$th component, and $\tilde{x}$ the reconstructed output.
    \item \textbf{Kullback-Leibler (KL) divergence:}
    \begin{equation}
    \begin{split}
        L_\mathrm{KL} &= D_\mathrm{KL}(p||q) = \int_x p(x)\ln \frac{p(x)}{q(x)} \\
        &= \frac{1}{2}\left[\mu^T\mu + \mathrm{Tr}(\Sigma) - L - \ln|\Sigma|\right],
    \end{split}
    \end{equation}
    where $L$ is the dimension of the latent space.
\end{enumerate}
The total loss is $L_\mathrm{tot} = L_\mathrm{recon} + L_\mathrm{KL}$.

The architecture used in this work consists of a fully connected (FC) layer with 2048 nodes, followed by a convolutional layer operating on $32 \times 64$ reshaped 2D images (with 4 channels, totaling 8192 dimensions), and another FC layer with 4096 nodes before entering the latent space. The decoder mirrors this structure, with the final output layer producing $2L$ values (mean and variance for each latent dimension). The total number of model parameters depends on the input size and can reach up to 600 million for the largest systems studied.

All neural network models are implemented in PyTorch, with a learning rate of $10^{-3}$ and weight decay of $10^{-2}$ for regularization. Weight decay is important to prevent overfitting, especially given the large model size. To ensure robust training, multiple runs with different random seeds are performed; for the main results, 100 independent models were trained and the best is selected, while 10 trials were used for other cases. The results are generally insensitive to initialization, except for rare outliers. Training was performed on NVIDIA H100 GPUs, with each run (100 epochs) taking approximately 1 hour for largest systems with $10$ trials.

Alternative architectures, including deeper fully connected networks and transformer-based models \cite{NIPS2017_3f5ee243}, were also tested (see SI \footnote{SI Sec. IV: VAE other architectures and comparison}), but the convolutional architecture described above consistently yielded the best performance, in terms of total loss, for form factor data.

\subsection{Principal Component Analysis} \label{med:PCA}

Principal component analysis (PCA) is used to identify the principal directions of variance in the form factor dataset, providing an efficient linear basis for dimensionality reduction. Given $N$ training samples, each represented as a complex vector $\mathbf{x}_n$ of dimension $D$, we first center the data by subtracting the mean $\bar{\mathbf{x}} = \frac{1}{N}\sum_{n}\mathbf{x}_n$. The data matrix $\mathbf{X}$ is constructed such that its $n$th row is $(\mathbf{x}_n - \bar{\mathbf{x}})^\dagger$, where the dagger denotes Hermitian conjugation to account for the complex nature of the data (alternative approach separating real and imaginary parts can be find in Ref. \cite{hellings2015composite}, which is beyond the scope of this work).

The covariance matrix is defined as
\begin{equation}
    \mathbf{S} = \frac{1}{N} \mathbf{X}^\dagger \mathbf{X},
\end{equation}
and the principal components $\mathbf{u}_j$ are obtained by solving the eigenvalue problem
\begin{equation}
    \mathbf{S} \mathbf{u}_j = \lambda_j \mathbf{u}_j,
\end{equation}
where $\lambda_j$ are the eigenvalues, ordered in descending order. Since the data dimension $D$ is typically much larger than the number of samples $N$, it is computationally efficient to solve the eigenproblem in the reduced $N \times N$ space:
\begin{equation}
    \tilde{\mathbf{S}} \mathbf{v}_j = \lambda_j \mathbf{v}_j, \quad \text{where} \quad \tilde{\mathbf{S}} = \frac{1}{N} \mathbf{X} \mathbf{X}^\dagger.
\end{equation}
The principal components in the original space are then reconstructed as
\begin{equation}
    \mathbf{u}_j = \frac{1}{\sqrt{N \lambda_j}} \mathbf{X}^\dagger \mathbf{v}_j,
\end{equation}
with normalization $||\mathbf{u}_j|| = 1$.

Each centered data point can be expanded in terms of the first $M$ principal components as
\begin{equation}
    \tilde{\mathbf{x}}_n - \bar{\mathbf{x}} = \sum_{j=1}^M a_{nj} \mathbf{u}_j,
\end{equation}
where the expansion coefficients are $a_{nj} = (\mathbf{x}_n - \bar{\mathbf{x}})^\dagger \mathbf{u}_j$. Truncating to the leading $M$ components provides an efficient low-rank approximation of the original data, capturing the dominant modes of variance.

This approach allows us to analyze the structure of the form factor dataset, identify the most relevant features, and efficiently reconstruct or generate new form factors by sampling in the principal component space.

\section{Acknowledgements}

We thank Jie Wang and Adrian Del Maestro for insightful discussions. The work at the UT Knoxville was primarily supported by the National Science Foundation Materials Research Science and Engineering Center program through the UT Knoxville Center for Advanced Materials and Manufacturing (DMR-2309083). The work at LANL was carried out under the auspices of the U.S. DOE NNSA under contract No. 89233218CNA000001 through the LDRD Program, and was supported by the Center for Nonlinear Studies at LANL, and was performed, in part, at the Center for Integrated Nanotechnologies, an Office of Science User Facility operated for the U.S. DOE Office of Science, under user proposals $\#2018BU0010$ and $\#2018BU0083$. KS was supported in part by Air Force Office of Scientific Research MURI FA9550-23-1-0334 and the Office of Naval Research MURI N00014-20-1-2479 and Award N00014-21-1-2770, and by the Gordon and Betty Moore Foundation Award N031710.  The computation for this work was performed on the University of Tennessee Infrastructure for Scientific Applications and Advanced Computing (ISAAC) computational resources.

\section{Author contributions statement}

A.-K. Wu and S.-Z. Lin designed the research. A.-K. Wu performed the calculations. L. Primeau and J. Zhang provided machine-learning and coding support. K. Sun and Y. Zhang contributed to the analysis and interpretation of the results. All authors discussed the data, revised the manuscript, and read and approved the final manuscript.

\section{Additional information}

\textbf{Data availability:}
The code and training models used in this study are publicly available at \url{https://github.com/angkunwu/pyMLFCI}. The datasets generated and/or analyzed during the current study are not publicly available due to their large size, but are available from the corresponding author on reasonable request.

\textbf{Competing interests:} The authors declare no competing financial or non-financial interests.


\bibliography{refs}

\end{document}



\title{Supplementary Information: Modeling Quantum Geometry for Fractional Chern Insulators with unsupervised learning}

\author{Ang-Kun Wu}
\affiliation{Department of Physics and Astronomy, University of Tennessee, Knoxville, Knoxville, Tennessee, 37996, USA}
\author{Louis Primeau}
\affiliation{Department of Physics and Astronomy, University of Tennessee, Knoxville, Knoxville, Tennessee, 37996, USA}
\author{Jingtao Zhang}
\affiliation{Google, Mountain View, CA, 94043, USA}
\author{Kai Sun}
\affiliation{Department of Physics, University of Michigan, Ann Arbor, Michigan, 48109, USA}
\author{Yang Zhang}
\affiliation{Department of Physics and Astronomy, University of Tennessee, Knoxville, Knoxville, Tennessee, 37996, USA}
\author{Shi-Zeng Lin}
\affiliation{Theoretical Division, T-4 and CNLS, Los Alamos National Laboratory (LANL),
Los Alamos, New Mexico 87545, USA}
\affiliation{Center for Integrated Nanotechnology, Los Alamos National Laboratory (LANL),
Los Alamos, New Mexico 87545, USA}

\date{\today}

\maketitle

\section{The original single-particle Hamiltonian and training form factors}

The single-particle Hamiltonian used to generate the training data is based on the quadratic band crossing point (QBCP) model—a low-energy two-band system with quadratic dispersion and a moir\'e periodic potential~\cite{PhysRevLett.130.216401}. The Hamiltonian is given by
\begin{equation}
    \begin{split}
        H_0(\mathbf{r}) &=  H_\Gamma(\mathbf{k}) + A_x(\mathbf{r})\sigma_x + A_y(\mathbf{r})\sigma_y, \\
        H_\Gamma(\mathbf{k}) &= (c_0 k^2 - m_z)\sigma_0 + 4(k_x^2 - k_y^2)\sigma_x + 8k_x k_y \sigma_y, \\
        \tilde{A}(\mathbf{r}) &= \frac{\alpha^2}{2} \sum_{n=1}^3 \omega^{n-1}\cos(\mathbf{G}_n \cdot \mathbf{r} + \phi),
    \end{split}
\end{equation}
where $\sigma_\alpha$ are the Pauli matrices and $\sigma_0$ is the identity. $H_\Gamma(\mathbf{k})$ describes the QBCP as a low-energy theory near the $\Gamma$ point of a kagome lattice. In this context, next-nearest-neighbor hopping tunes the parameter $c_0$. An additional phase in the nearest-neighbor hopping introduces $m_z$, breaking time-reversal symmetry~\cite{PhysRevLett.135.016501}. The periodic strain $\tilde{A}(\mathbf{r})$ is modeled using the first-harmonic approximation, where $\alpha$ is the strain strength, $\omega = e^{2\pi i/3}$, and the reciprocal lattice vectors are $\mathbf{G}_1 = \frac{4\pi}{\sqrt{3}}(0,1)$, $\mathbf{G}_2 = \frac{4\pi}{\sqrt{3}}(-\sqrt{3}/2, -1/2)$, and $\mathbf{G}_3 = \frac{4\pi}{\sqrt{3}}(\sqrt{3}/2, -1/2)$.

In the chiral limit ($c_0 = 0$), the model exhibits exactly flat topological bands at $E = 0$ with Chern number $C = \pm 1$ and ideal quantum metric at the so-called \textbf{magic dimensionless parameters} $\tilde{\alpha}_j = \alpha/|\mathbf{G}^m| = 0.78943,\, 2.1325,\, 3.517548,\,\ldots$. Here, the flat Chern bands are sublattice-polarized, defined as eigenstates of $\sigma_z$. Introducing a finite $m_z$ breaks time-reversal symmetry, splitting the energies of the two Chern bands; the sign of $m_z$ determines the sign of the Hall conductance in the noninteracting regime. In the chiral limit, we set $m_z = 0$ and use sublattice polarization to separate the degenerate flat bands at the band center.

Beyond the chiral limit, the band dispersion and quantum metric of the two low-energy topological flat bands can be tuned by $c_0$, which modifies the Berry curvature and violates the trace condition. In this regime, we set $m_z = 1.0$ to separate the two Chern bands. Notably, at the chiral magic parameters ($c_0 = 0$, $\tilde{\alpha}_j$), the flatband form factor is identical whether obtained via sublattice polarization or by breaking time-reversal symmetry ($m_z = 1.0$). However, $m_z$ modifies the bands away from the magic parameters, which is why sublattice polarization is used to collect data in the chiral case.

Overall, this flatband model has two key tunable parameters: $H_0 = H_0(\tilde{\alpha}, c_0)$. Training data are generated by varying $\tilde{\alpha} \in [0.35, 3.55]$ and $c_0 \in [-1.0, 1.0]$ at each of the first three magic parameters. The system sizes for the training data are $N_x \times N_y = 4 \times 6$ in the main text (using a uniform mesh along directions $\mathbf{G}_x = \mathbf{G}_2$, $\mathbf{G}_y = \mathbf{G}_3$), and the form factors are calculated from the eigenstates of the lower middle Chern band.

\begin{figure}[t!]
\begin{center}
\setlabel{pos=nw,fontsize=\large,labelbox=false}
\xincludegraphics[scale=0.27,label=a]{FCINx4Ny6a2.pdf}
\xincludegraphics[scale=0.27,label=b]{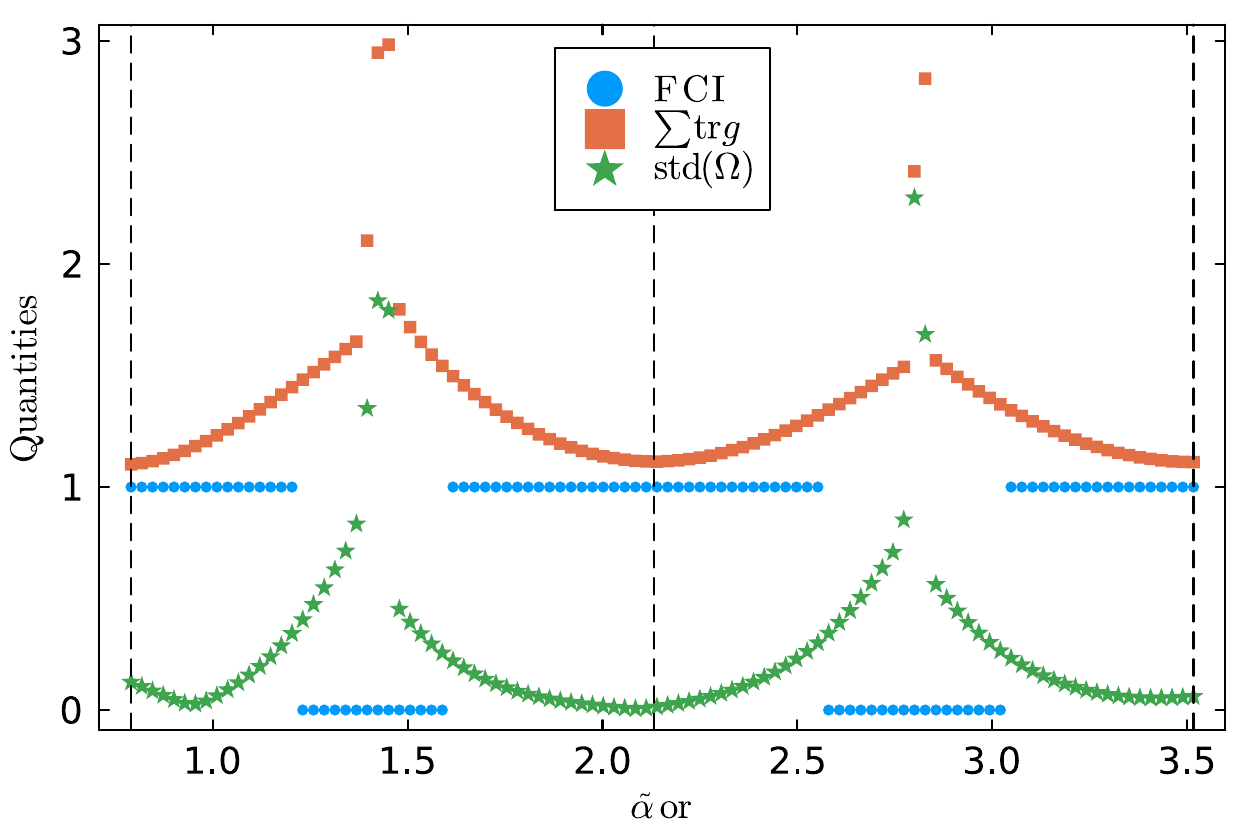}
\xincludegraphics[scale=0.27,label=c]{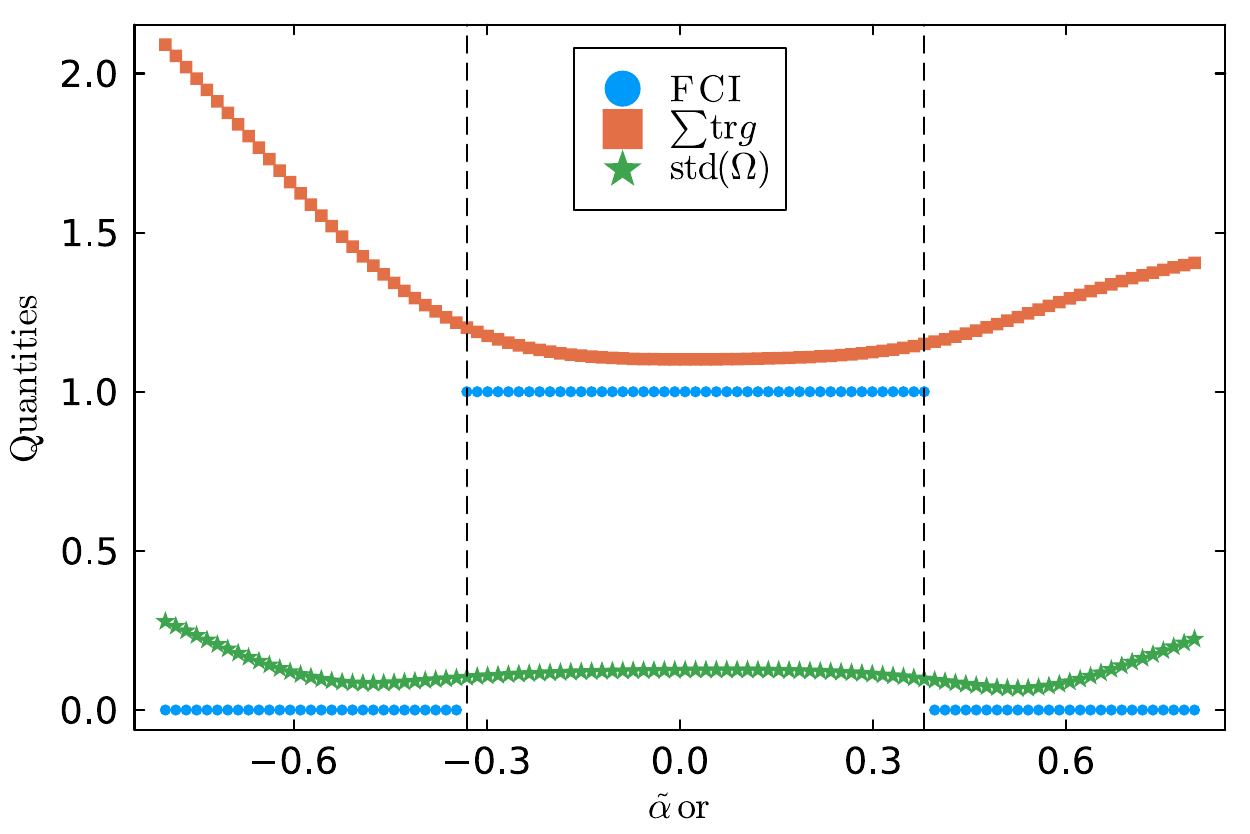}
\caption{\label{sifig:traindata}
Training data and many-body diagnostics across the model parameter space. (a) Many-body energy spectrum of the fractional Chern insulator (FCI) state at the second magic parameter $\tilde{\alpha}_2 = 2.1325$ with $c_0=0$, $N_x=4$, and $N_y=6$. The inset shows the particle entanglement spectrum (PES) with a gap at the $(1,3)$-admissible counting for $Q_A=3$ at filling factor $\nu=1/3$ and $Q=8$. (b) Many-body phase labels and discrete quantum geometry of the form factors as functions of $\tilde{\alpha}$. FCI states are labeled by 1 and non-FCI states are labeled by 0. The system size is $N_x=3$, $N_y=5$ for visualization. Dashed vertical lines mark the magic-parameter positions. (c) Many-body phase labels and discrete quantum geometry of the form factors as functions of $c_0$ at $\tilde{\alpha}_2$ for $N_x=3$, $N_y=5$. Dashed vertical lines mark the transition points at $c_0=-0.33$ and $0.38$.
}
\end{center}
\end{figure}

The standard FCI many-body energy spectrum at the second magic parameter $\tilde{\alpha}_2$ is shown in Fig.~\ref{sifig:traindata}(a). The gap in the particle entanglement spectrum (PES), together with ground state degeneracy at specific momentum sectors, provides evidence for the FCI nature of the ground states~\cite{PhysRevX.1.021014}. The PES calculation follows Ref.~\cite{PhysRevResearch.6.L032063}. The structure of the training form factor data across $\tilde{\alpha}$ and $c_0$ is shown in Fig.~\ref{sifig:traindata}(b) and (c), where data are collected around three magic parameters.

\begin{figure}[h!]
\begin{center}
\setlabel{pos=nw,fontsize=\large,labelbox=false}
\xincludegraphics[scale=0.04,label=a]{FFreala2Nx4Ny6.pdf}
\xincludegraphics[scale=0.04,label=b]{FFimaga2Nx4Ny6.pdf}
\xincludegraphics[scale=0.04,label=c]{FFabsa2Nx4Ny6.pdf}
\xincludegraphics[scale=0.04,label=d]{FFphasea2Nx4Ny6.pdf}
\caption{\label{sifig:ff}
Form factors in the central Brillouin zone illustrate the data representation used for machine learning. The form factors are shown in the central Brillouin zone (BZ) at $\tilde{\alpha}_2=2.1325$ with $c_0=0$, $N_x=4$, and $N_y=6$. (a) Real part of the form factor. (b) Imaginary part of the form factor. (c) Amplitude of the form factor. (d) Phase of the form factor.
}
\end{center}
\end{figure}

In Fig.~\ref{sifig:ff}, we present various visualizations of the form factors in the central Brillouin zone (BZ). It is important to note that the form factors are gauge-dependent, so their appearance will vary depending on the gauge choice. For example, there is a block at $k_x=3$ that appears purely real due to the current gauge. Physically, our understanding of the form factors is not based on their visual patterns. The form factor encodes the Hilbert space structure both locally and globally. Locally, the form factor is related to the gauge-invariant quantum geometry tensor, whose real part is the quantum metric and imaginary part is the Berry curvature. However, for machine learning applications, representing the complex form factor in terms of phase and amplitude leads to poor performance due to nonlinearity. Therefore, we use the real and imaginary parts of the form factors, which connects more directly to the physical quantities like Berry curvature and quantum metric, as the training data.

\section{Discrete Quantum geometry from two calculations}

\begin{figure}[t!]
\begin{center}
\setlabel{pos=nw,fontsize=\large,labelbox=false}
\xincludegraphics[scale=0.37,label=a]{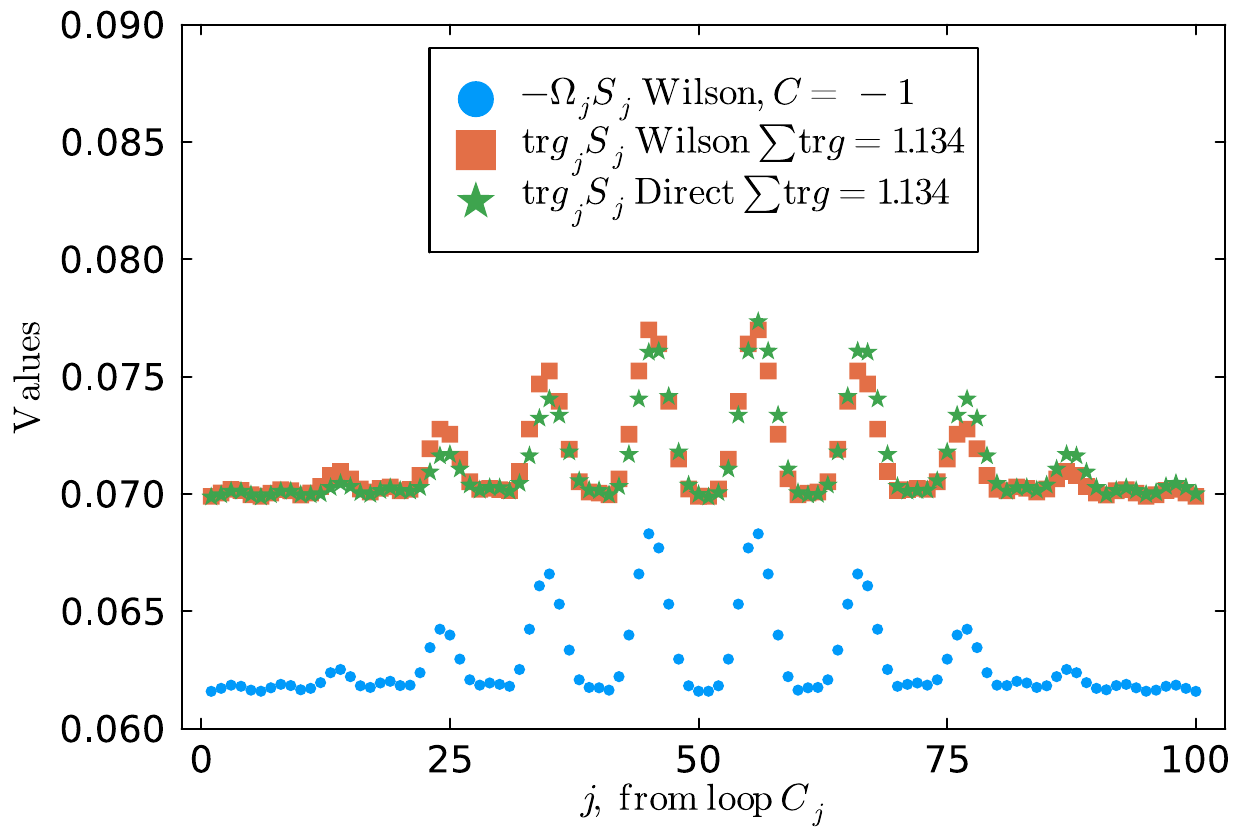}
\xincludegraphics[scale=0.37,label=b]{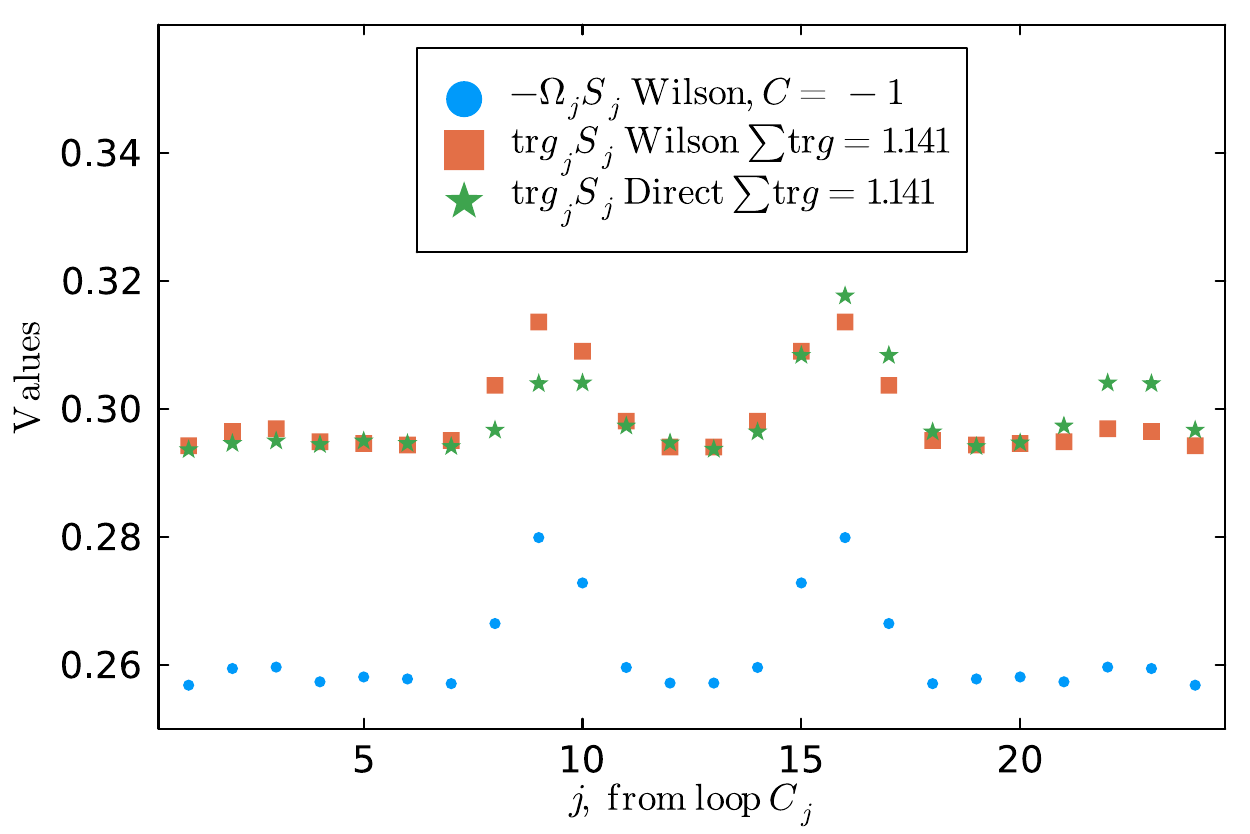}
\caption{\label{sifig:geometry}
Discrete Berry curvature and quantum metric benchmark the geometry extracted from finite momentum-space meshes. The figure compares the discrete Berry-curvature contribution $-\Omega_j S_j$ and the trace of the quantum metric multiplied by loop area, $\mathrm{tr}(g_j)S_j$, at the second magic parameter $\tilde{\alpha}_2$. The legend reports the Chern number and the total quantum metric. (a) Discrete quantum geometry for the $N_x=10$, $N_y=10$ system. (b) Discrete quantum geometry for the $N_x=4$, $N_y=6$ system, shown for point-to-point comparison with the machine-learning training data.
}
\end{center}
\end{figure}

To compare the machine-learned form factors with those obtained from the single-particle Hamiltonian, it is necessary to bridge the discrete form factors used in finite-size calculations with the continuum description of quantum geometry. In the continuum, the local curvature of Bloch wavefunctions is characterized by the quantum geometric tensor~\cite{provost1980riemannian}:
\begin{equation}
    \eta_{\mu\nu}(\mathbf{k}) \equiv \langle \partial_\mu u_\mathbf{k}|[1-|u_\mathbf{k}\rangle \langle u_\mathbf{k}|] |\partial_\nu u_\mathbf{k}\rangle,
\end{equation}
where $|u_\mathbf{k}\rangle$ is the Bloch eigenstate at momentum $\mathbf{k}$. The imaginary part of $\eta_{\mu\nu}$ yields the Berry curvature, $\Omega(\mathbf{k}) = -2\Im [\eta_{xy}(\mathbf{k})]$, while the real part gives the Fubini-Study metric, $g_{\mu\nu}(\mathbf{k}) = \Re[\eta_{\mu\nu}(\mathbf{k})]$. The Berry curvature provides a lower bound for the quantum metric, $\mathrm{tr}\,g(\mathbf{k}) = g_{xx}(\mathbf{k}) + g_{yy}(\mathbf{k}) \geq |\Omega(\mathbf{k})|$, with the ideal quantum geometry saturating this bound at all $\mathbf{k}$.

A straightforward way to estimate the quantum geometric tensor is to numerically evaluate the derivatives $|\partial_\mu u_\mathbf{k}\rangle$ and compute $\eta_{\mu\nu}(\mathbf{k})$ directly. However, this approach is not gauge-invariant and can be numerically unstable, especially on a discrete $\mathbf{k}$-mesh. To address this, we use a gauge-invariant and numerically robust Wilson loop approach, which is also more closely related to the structure of the form factors.

For the Berry curvature, recall that $\Omega(\mathbf{k}) = i(\langle \partial_x u|\partial_y u\rangle - \langle \partial_y u|\partial_x u\rangle) = \nabla_\mathbf{k} \times \mathbf{A}$, where the Berry connection is $\mathbf{A} = i\langle u|\nabla_\mathbf{k}|u\rangle$. On a discrete mesh, the Berry phase accumulated around a closed loop $C_j$ in momentum space can be computed as
\begin{equation}
    \phi_j = \mathrm{Im} \log \prod_{\mathbf{k}_l \in C_j} \langle u_{\mathbf{k}_l}|u_{\mathbf{k}_{l+1}}\rangle,
\end{equation}
where the product is taken around the loop. The Berry curvature at the central momentum $\mathbf{k}_j$ of the loop is then approximated by
\begin{equation}
    \Omega(\mathbf{k}_j) S_j \approx -\phi_j,
\end{equation}
where $S_j$ is the area enclosed by the loop. Summing over all loops in the Brillouin zone yields the total Chern number,
\begin{equation}
    C = \frac{1}{2\pi} \sum_j \Omega(\mathbf{k}_j) S_j = -\frac{1}{2\pi} \sum_j \phi_j.
\end{equation}
The sign convention depends on the orientation of the loops (here, we use clockwise).

Similarly, the Fubini-Study quantum metric can be evaluated from the squared overlap of neighboring Bloch states:
\begin{equation}
    s^2(\mathbf{k},\mathbf{k+q}) = 1 - |\langle u_\mathbf{k}|u_{\mathbf{k+q}}\rangle|^2 \underset{q\to 0}{\approx} g_{\mu\nu} q_\mu q_\nu,
\end{equation}
where $s^2$ is the quantum distance. For a simple square Wilson loop, the local trace of the metric can be directly estimated via two-point estimation
\begin{equation}
    \mathrm{tr}[g(\mathbf{k}_j)] S_j \approx \sum_{\mu=x,y} \frac{s^2(\mathbf{k}_j,\mathbf{k}_j + d\mathbf{q}_\mu) + s^2(\mathbf{k}_j,\mathbf{k}_j - d\mathbf{q}_\mu)}{2},
\end{equation}
where $d\mathbf{q}_\mu$ are the discrete steps in the $x$ and $y$ directions. Alternatively, the metric can be estimated by averaging the quantum distances along the edges of the Wilson loop (a four-point estimator similar to the Berry curvature estimation) as shown in the method of the main text, Eq. (10).

Numerically, both the Wilson loop (main text) and the direct estimation method yield similar results for the total quantum metric $\sum_j \mathrm{tr}[g(\mathbf{k}_j)] S_j$, though the Wilson loop approach more closely tracks the detailed behavior of the Berry curvature (see Fig.~\ref{sifig:geometry}). Unlike the Chern number, the total metric is more sensitive to finite-size effects due to the discretization of $\mathbf{q}$. We denote the normalized total metric as $\sum \mathrm{tr} g \equiv \sum_j \mathrm{tr}[g(\mathbf{k}_j)] S_j /(2\pi)$, which approaches the absolute value of the Chern number $|C|$ in the ideal limit.

Figure~\ref{sifig:geometry} benchmarks the discrete quantum geometry at the second magic parameter $\tilde{\alpha}_2$ for two system sizes. The Chern number and total quantum metric are shown in the legend. Panel (a) shows results for a large system ($N_x=10$, $N_y=10$), while panel (b) shows a smaller system ($N_x=4$, $N_y=6$) for direct comparison with the machine learning training data. The agreement between the two calculations demonstrates the reliability of the discrete geometry extraction, even for relatively small system sizes.

This detailed connection between discrete form factors and continuum quantum geometry provides a robust framework for benchmarking machine-learned form factors and understanding their physical content.

\section{Classification of the training data}

Before exploring unsupervised learning, we first perform supervised classification \cite{bishop2006pattern} of the training data to test the ability of machine learning models to distinguish FCI and non-FCI form factors. The labeled dataset consists of 400 samples of form factors with $N_x=3$, $N_y=5$, each labeled as either FCI or non-FCI. We employ three supervised learning approaches using the \texttt{scikit-learn} Python package~\cite{scikit-learn}: logistic regression (LR, a linear model), a fully connected neural network with two hidden layers of size $[50,50]$ (FCNN, a nonlinear model), and a support vector machine with linear kernel (SVM, a nonlinear model). Model performance is evaluated using 5-fold cross-validation to avoid overfitting.

The $400$ training samples are drawn from four parameter regimes: (a) $100$ samples from $\tilde{\alpha}\in [\tilde{\alpha}_1, \tilde{\alpha}_3]$ with $c_0=0$ (chiral limit, see Fig.~\ref{sifig:traindata}b); (b) $100$ samples from $c_0\in [-0.8,0.8]$ at $\tilde{\alpha}=\tilde{\alpha}_1$, $m_z=1$; (c) $100$ samples from $c_0\in [-0.6,0.6]$ at $\tilde{\alpha}=\tilde{\alpha}_2$, $m_z=1$ (see Fig.~\ref{sifig:traindata}c); and (d) $100$ samples from $c_0\in [-0.5,0.5]$ at $\tilde{\alpha}=\tilde{\alpha}_3$, $m_z=1$.

The cross-validation accuracies are summarized in Table~\ref{sitab:classification}. The linear model (LR) achieves an average accuracy of $86\%$, while the nonlinear models (FCNN and SVM) yield higher accuracies of $94\%$ and $95\%$, respectively. The FCNN results remain robust when increasing the number of nodes or layers. These results indicate that the form factors can be effectively classified using machine learning. While the FCNN approaches the SVM accuracy, it requires more computational resources. Notably, although the linear model does not achieve very high accuracy, the SVM with a linear kernel performs well, suggesting that the data are linearly separable in a higher-dimensional space, which motivates the unsupervised learning approaches on the complex form factors in the main text.

\begin{table}[h]\label{sitab:classification}
    \centering
    \begin{tabular}{|c|c|c|c|}
        \hline
        Method & Logistic Regression & Fully Connected Neural Network & Support Vector Machine \\
        \hline
        Max Accuracy & 92\% & 97\% & 97\% \\
        Min Accuracy & 81\% & 92\% & 92\% \\
        Average Accuracy & 86\% & 94\% & 95\% \\
        \hline
    \end{tabular}
    \caption{
    Classification accuracy benchmarks show that nonlinear models outperform logistic regression on the form-factor dataset. The table summarizes the classification performance for 400 form-factor samples with $N_x=3$ and $N_y=5$. Accuracy is evaluated by five-fold cross-validation. LR denotes logistic regression, FCNN denotes a fully connected neural network, and SVM denotes a support vector machine. Max, Min, and Average report the maximum, minimum, and mean classification accuracies, respectively, across the cross-validation splits.}
\end{table}

\section{Other VAE architectures and comparison}

In addition to the VAE architecture in the main text  (a single convolutional layer sandwiched between two fully connected layers), we explored several alternative architectures to assess performance and robustness. The architectures tested include: (1) fully connected neural networks (FCNN) with four hidden layers; (2) a convolutional layer followed by two fully connected layers; (3) a convolutional layer plus one transformer layer, then fully connected layers; (4) the architecture in the main text; and (5) a transformer layer sandwiched between two fully connected layers.

Directly applying a convolutional layer to the input form factors performs poorly, yielding featureless results. This may be due to the translational symmetry across the Brillouin zones (BZs): in the uniform Berry curvature limit, the form factors are locally similar, making it difficult for convolutional filters to extract meaningful features. Similarly, transformer-based architectures also underperform, likely because there are no ``important tokens'' in the form factors across the BZ.

Table~\ref{sitab:architecture} compares the performance of three representative architectures that performed reasonably well even for small datasets: (1) FCNN with four hidden layers ($[2048, 1024, 512, 256]$); (2) the ``sandwiched convolution'' architecture in the main text; and (3) a transformer layer sandwiched between two fully connected layers (``sandwiched transformer''). For FCNN and convolutional architectures, we varied the latent space dimension ($2, 3, 4$) and, for the convolutional architecture, the kernel size ($3, 5$). All results are for the FCI form factors with $N_x=3$, $N_y=5$, as used in the classification task. Note that the loss values are subject to randomness from the latent space sampling inherent in VAE training.

All three architectures yield reasonable results (can give generated FCI form factors) with controlled reconstruction loss. While the total loss $L_\mathrm{tot}$ is similar across models, the reconstruction loss $L_\mathrm{recon}$ (sum of squared errors) is most indicative of the quality of the learned form factors. Even though the reconstruction losses are comparable, the quality of the reconstructed form factors, as assessed by exact diagonalization (ED) spectra, can vary significantly. Alternative loss functions to the total squared error could be explored in future work. Overall, the sandwiched convolution architecture consistently gives the best results, both in terms of loss and when comparing the reconstructed form factors via ED spectra. The FCNN can also perform well, but its results are less robust to initialization and require larger hidden dimensions. The transformer-based architecture performs systematically worse, likely due to the lack of salient tokens in the form factors. Notably, a latent space dimension of $3$ gives the best results, consistent with the training data being generated around three magic parameters. A kernel size of $3$ outperforms $5$, suggesting that local structure is more important than global structure—consistent with the poor performance of transformer models. Overall, the sandwiched convolution architecture with latent dimension $3$ and kernel size $3$ provides the best balance of performance and robustness.

\begin{table}[h]\label{sitab:architecture}
    \centering
    \begin{tabular}{ |c|c|c|c|c|c| }
        \hline
        Architecture & Latent Dim & Kernel Size & Mean $L_\mathrm{tot}$ &Mean $L_\mathrm{recon}$ & Mean $L_\mathrm{KL}$ \\
        \hline
        FCNN & 2 & N/A & 4.16 & 0.47 & 3.69 \\
        FCNN & 3 & N/A & 4.40 & 0.59 & 3.81 \\
        FCNN & 4 & N/A & 4.24 & 0.53 & 3.72 \\
        \hline
        Sandwiched Convolution & 2 & 3 & 4.41 & 0.38 & 4.03 \\
        {\color{red}Sandwiched Convolution} & {\color{red} 3} & {\color{red} 3} & {\color{red} 3.83} & {\color{red} 0.34} & {\color{red} 3.50} \\
        Sandwiched Convolution & 4 & 3 & 4.36 & 0.35 & 4.01 \\
        Sandwiched Convolution & 3 & 5 & 4.97 & 0.54 & 4.44 \\
        \hline
        Sandwiched Transformer & 3 & N/A & 4.62 & 1.57 & 3.05 \\
        \hline
    \end{tabular}
    \caption{
    Variational-autoencoder architecture comparisons identify the most effective model for reconstructing fractional-Chern-insulator form factors. The table compares variational autoencoder (VAE) architectures for form factors with $N_x=3$ and $N_y=5$ using only fractional Chern insulator (FCI) samples for training and loss evaluation. FCNN denotes a fully connected neural network. Kernel Size gives the convolutional kernel size when applicable, and N/A indicates that no convolutional kernel is used. Mean $L_\mathrm{tot}$, Mean $L_\mathrm{recon}$, and Mean $L_\mathrm{KL}$ denote the mean total loss, reconstruction loss, and Kullback-Leibler divergence loss, respectively. The architecture used in the main text is highlighted in red. All trained models are selected from 10 independent training runs.}
\end{table}

\begin{figure}[h!]
\begin{center}
\setlabel{pos=nw,fontsize=\large,labelbox=false}
\xincludegraphics[scale=0.28,label=a]{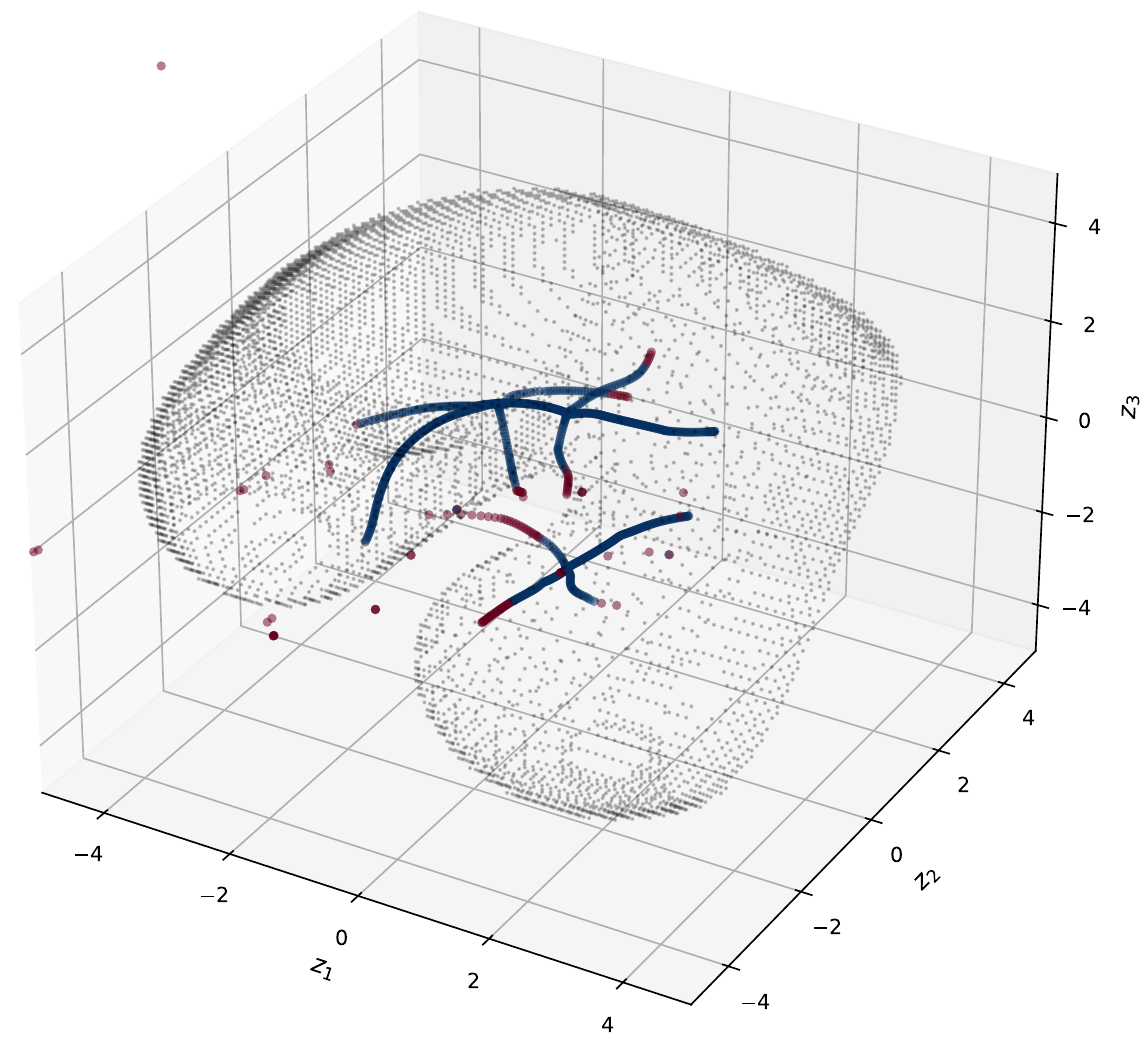}
\xincludegraphics[scale=0.36,label=b]{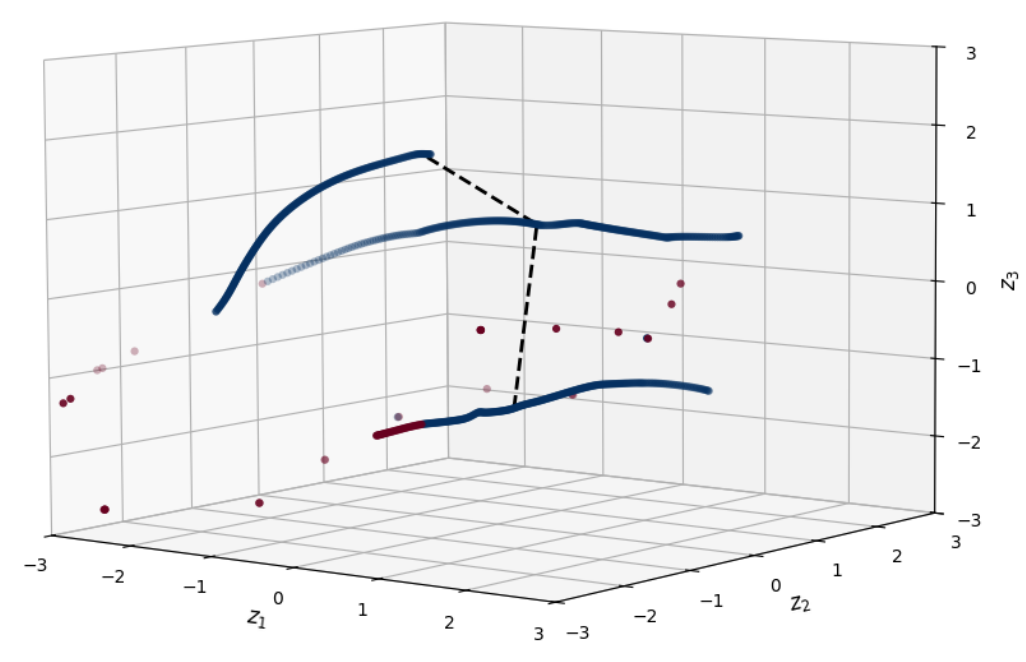}
\xincludegraphics[scale=0.29,label=c]{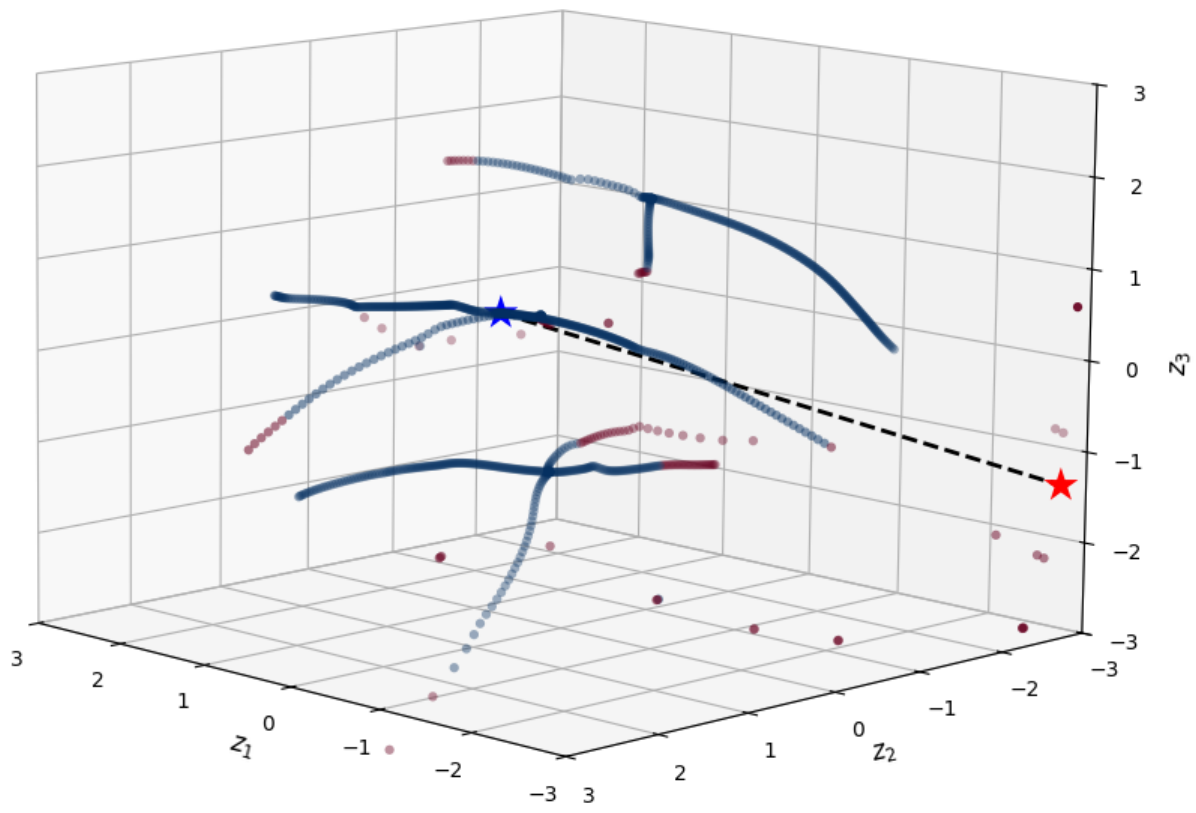}
\caption{\label{sifig:latent}
Latent-space structure reveals phase separation, interpolation paths, and charge-density-wave generation in the variational autoencoder. The figure shows details of the latent space corresponding to Fig.~2b in the main text. (a) Decision boundary determined by a support vector machine (SVM), shown as black dots, overlaid on the latent space of the training data. (b) Trajectory shown as a dashed line in latent space connecting the three magic-parameter points; here the training data include only cases with varying $\tilde{\alpha}$. (c) Trajectory shown as a dashed line in latent space connecting sampled charge density wave (CDW) points, marked by red stars, to the ideal fractional Chern insulator (FCI) at the second magic parameter, marked by a blue star, on top of the training-data latent space.
}
\end{center}
\end{figure}

Fig.~\ref{sifig:latent} illustrates the detailed structure of the latent space corresponding to the training data in Fig.~2b of the main text. As shown in Fig.~\ref{sifig:latent}a, applying a support vector machine (SVM) with a radial basis function kernel enables the identification of the decision boundary separating FCI form factors in the latent space. For $1000$ training samples with $N_x=4$, $N_y=6$, the SVM achieves an accuracy of $0.993$, demonstrating that the latent space exhibits clear phase separation under a nonlinear kernel. Notably, while a linear kernel suffices for classification in the original form factor space, the reduced dimensionality of the latent space necessitates a nonlinear approach for optimal separation. Fig.~\ref{sifig:latent}b presents the trajectories obtained by interpolating between the three ideal form factors at the magic parameters. In the single-particle model, varying $\tilde{\alpha}$ traces out three branches that extend toward distant non-FCI samples (red points), whereas the dashed line in the latent space provides a plausible interpolation connecting ideal form factors that are close in this representation. Finally, Fig.~\ref{sifig:latent}c shows the trajectory from a representative CDW form factor (red star) to the nearest ideal FCI form factor (blue star) within the latent space.

\begin{figure}[h!]
\begin{center}
\setlabel{pos=nw,fontsize=\large,labelbox=false}
\xincludegraphics[scale=0.30,label=a]{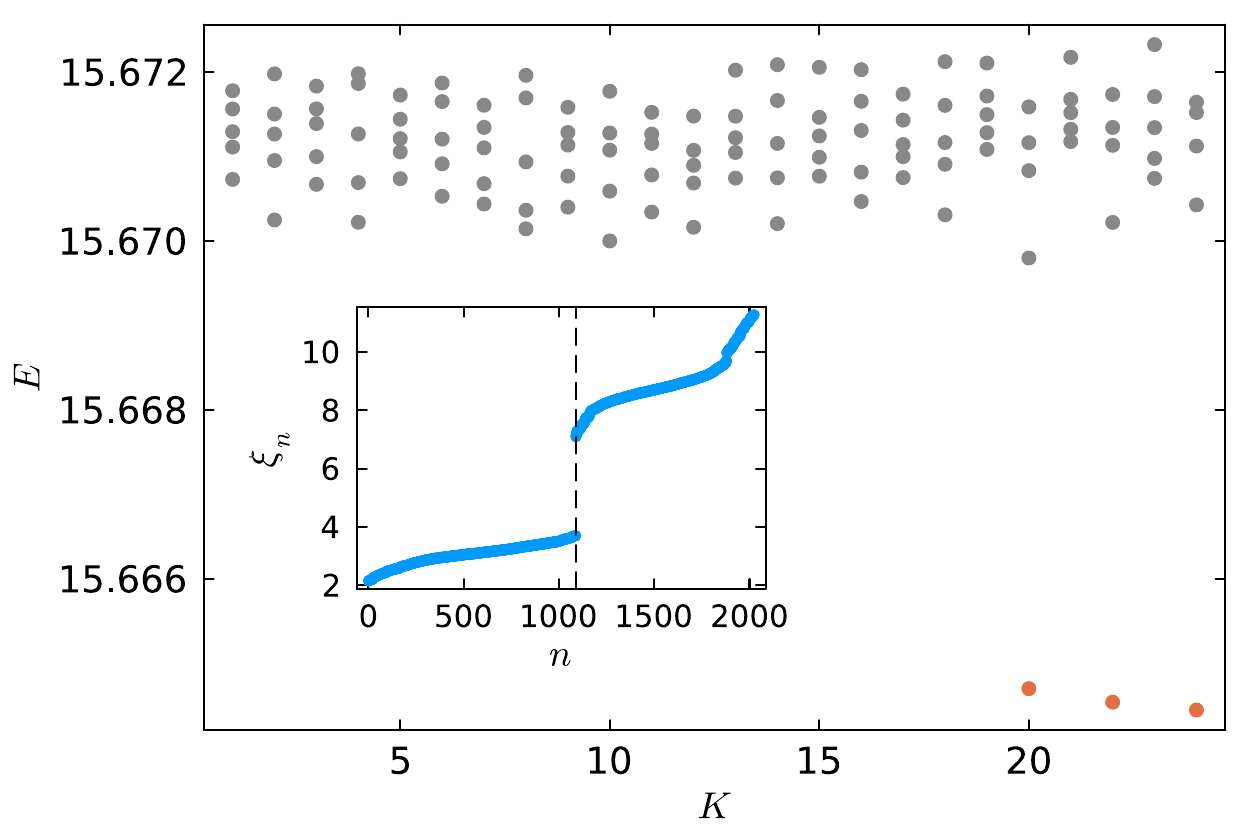}
\xincludegraphics[scale=0.30,label=b]{VAEFCiNx4Ny6FCICDWInd73.pdf}
\xincludegraphics[scale=0.30,label=c]{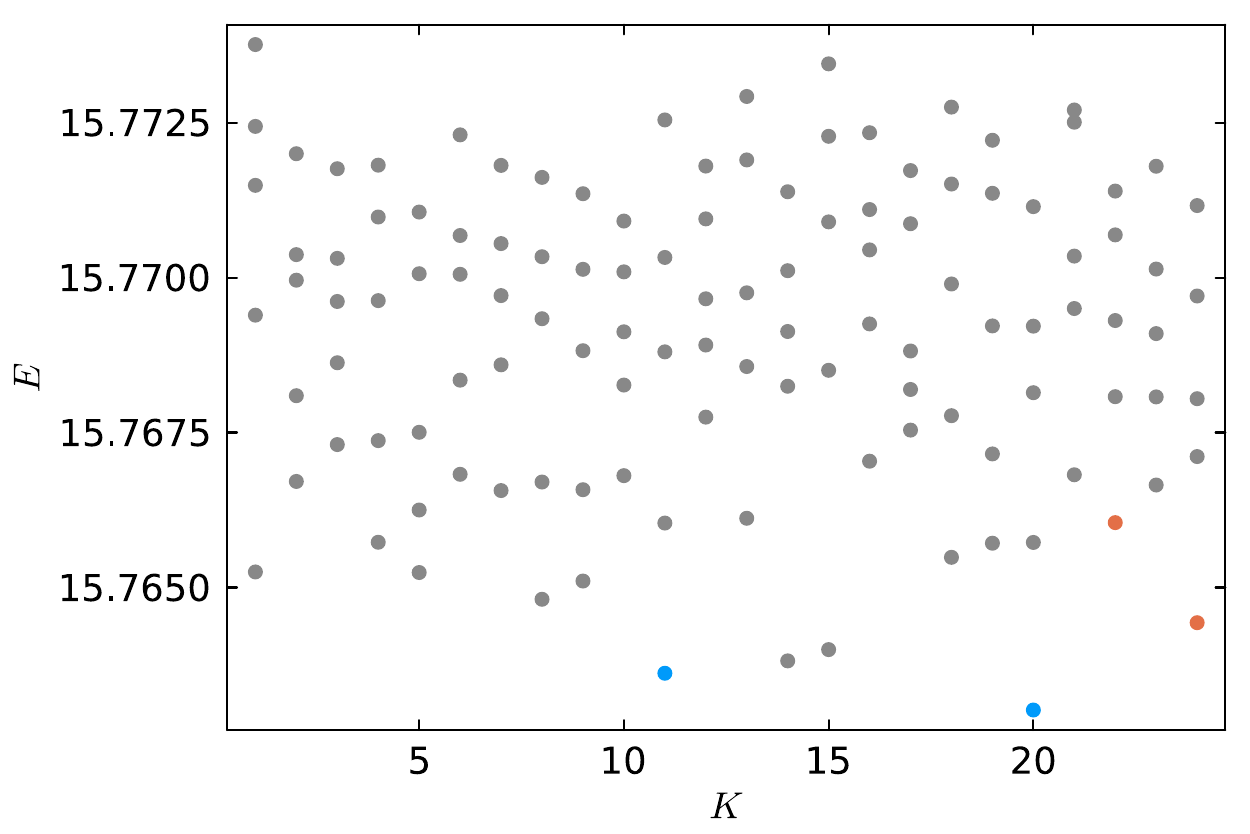}
\xincludegraphics[scale=0.30,label=d]{CDWVAENx4Ny6Ind100.pdf}
\caption{\label{sifig:VAEcross}
Many-body spectra evolve continuously from a fractional Chern insulator to a charge density wave along a variational-autoencoder path. The figure shows the evolution of the many-body energy spectrum along the variational autoencoder (VAE) latent-space path from a fractional Chern insulator (FCI) state to a charge density wave (CDW) state. (a) FCI spectrum at the latent-space coordinate $z$ corresponding to $\tilde{\alpha}_2$ with $N_x=4$ and $N_y=6$. (b) Spectrum at an intermediate state after leaving the FCI phase along the VAE path to the CDW state. (c) Spectrum at an intermediate state after entering the CDW phase along the same path. (d) CDW spectrum at an unseen state generated in the VAE latent space.
}
\end{center}
\end{figure}

Fig.~\ref{sifig:VAEcross} shows the evolution of the many-body spectrum from FCI to CDW along the latent space trajectory, as illustrated in Fig.~3c of the main text. The trajectory in latent space produces a smooth evolution of the spectrum: the gap closes at the intermediate state and reopens in the CDW phase. The ground state degeneracy changes from three-fold (three quasi-degenerate ground states for $\nu=1/3$) to twofold (two charge-ordered ground states) across the transition. This smooth evolution highlights the effectiveness of VAE latent space modeling for the form factors.

\section{PCA expansion and ED results}

Principal component analysis (PCA) provides a simple linear method to reduce the dimensionality of the form factor data according to the variance in the dataset~\cite{bishop2006pattern}. By projecting onto the leading principal components, we can approximate the form factors by truncating the less significant components:
\begin{equation}
    \tilde{\mathbf{x}}_n-\bar{\mathbf{x}} =\sum_{j=1}^M a_{nj} \mathbf{u}_j=\sum_{j=1}^M [(\mathbf{x}_n-\bar{\mathbf{x}})^\dagger \mathbf{u}_j] \mathbf{u}_j,
\end{equation}
where $\mathbf{x}_n$ is the $n$-th form factor, $\bar{\mathbf{x}}$ is the mean, $\mathbf{u}_j$ is the $j$-th principal component, $a_{nj}$ is the projection coefficient, and $M$ is the number of principal components retained. Here, we apply PCA only to FCI form factors; results for PCA on all training data (FCI and non-FCI) are similar in terms of geometry and variance spectrum (the center of the PCA, the mean data $\bar{\mathbf{x}} = \frac{1}{N}\sum_{n}\mathbf{x}_n$ will be different), but the reconstructed form factors differ due to different data means $\bar{\mathbf{x}}$.

\begin{figure}[t!]
\begin{center}
\setlabel{pos=nw,fontsize=\large,labelbox=false}
\xincludegraphics[scale=0.27,label=a]{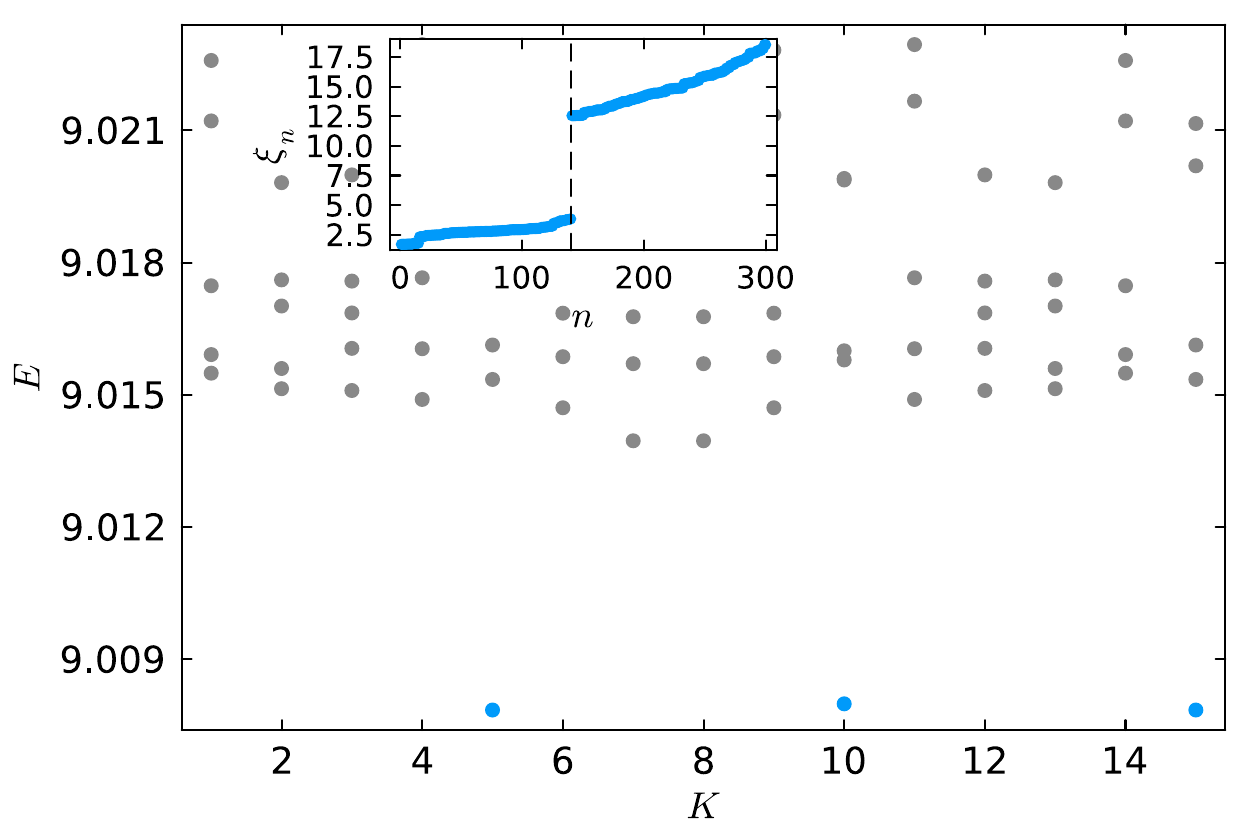}
\xincludegraphics[scale=0.27,label=b]{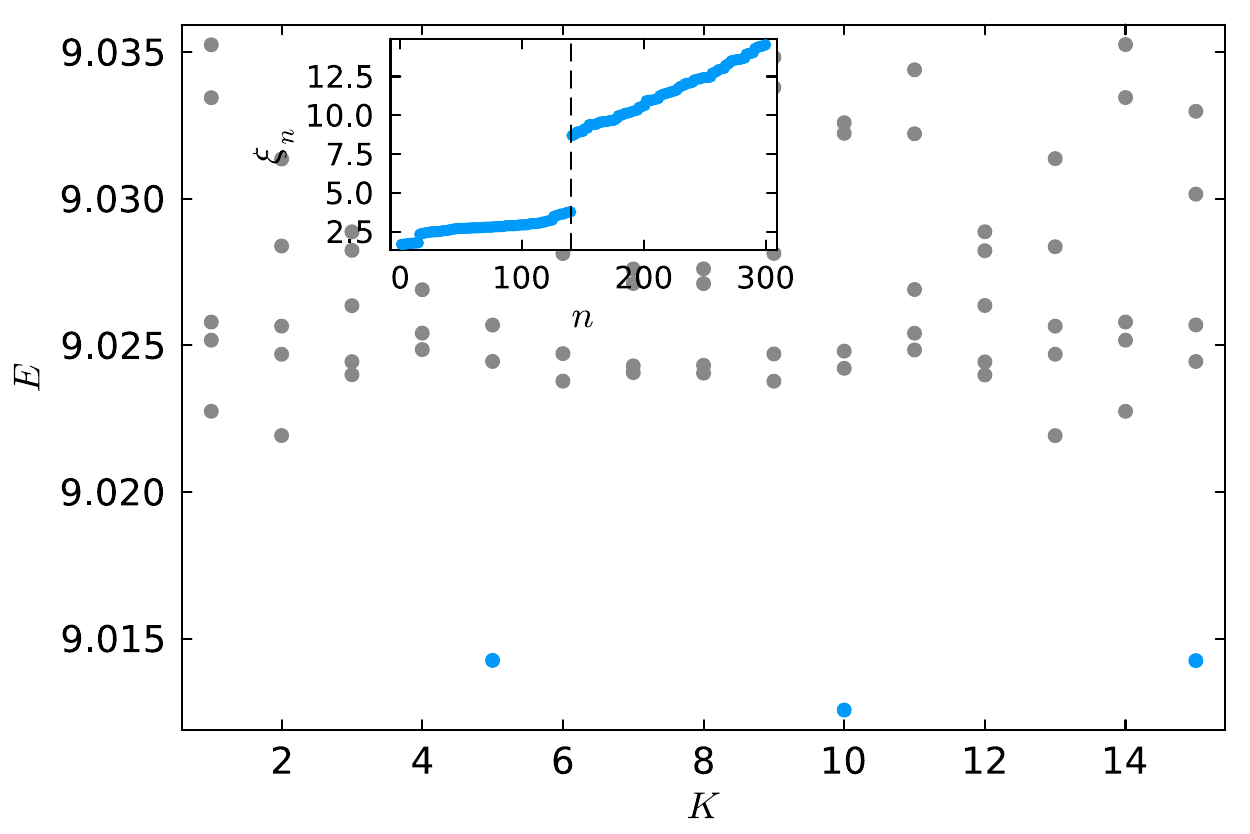}
\xincludegraphics[scale=0.27,label=c]{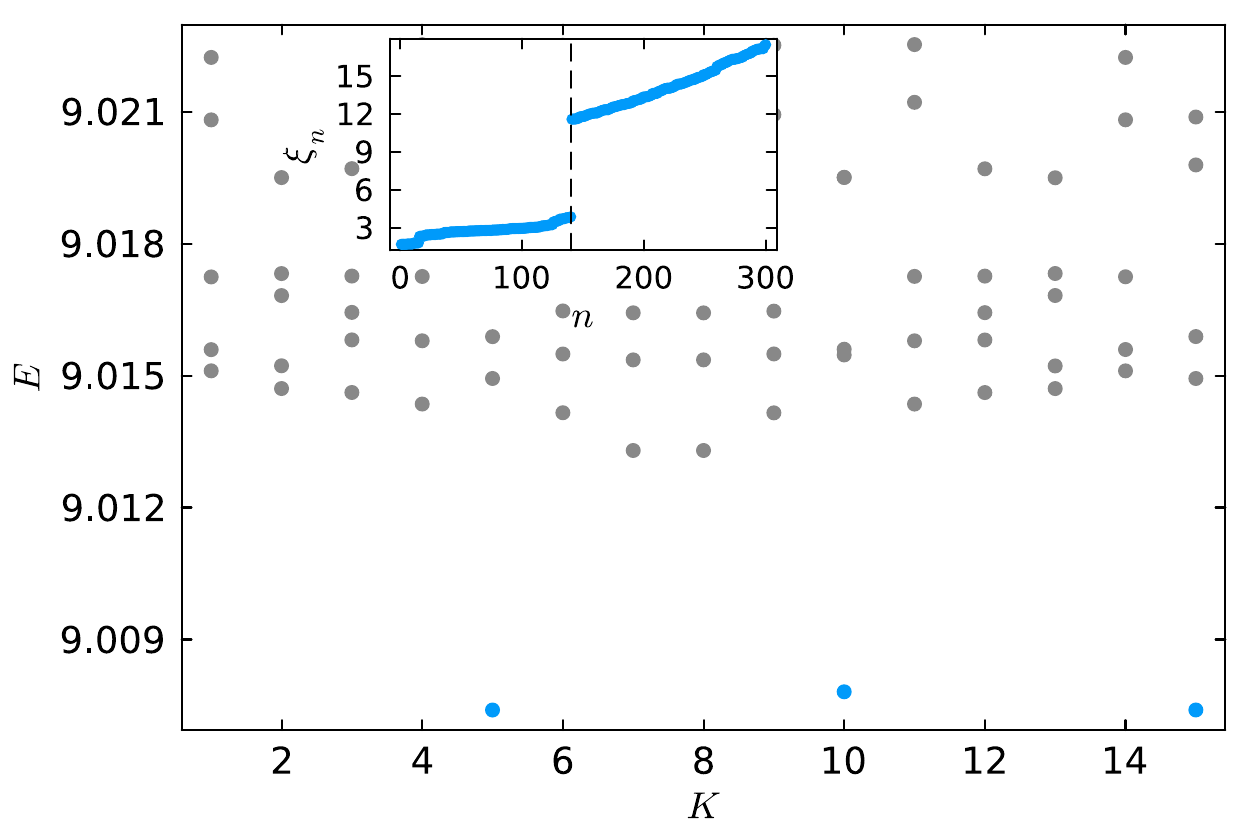}
\caption{\label{sifig:FCIPCA}
Principal-component truncation preserves the characteristic many-body spectrum of fractional Chern insulator form factors. The figure shows exact-diagonalization (ED) spectra for principal component analysis (PCA) truncations of fractional Chern insulator (FCI) form factors at $\tilde{\alpha}_1$ with $N_x=3$ and $N_y=5$. (a) Original form factor. (b) PCA-truncated form factor with $M=1$ retained principal component. (c) PCA-truncated form factor with $M=20$ retained principal components.
}
\end{center}
\end{figure}

To demonstrate the impact of PCA truncation on many-body spectra, we consider $N_x=3$, $N_y=5$ FCI form factors at $\tilde{\alpha}_1$ with $c_0=0$, where finite-size effects are more pronounced. The original form factor [Fig.~\ref{sifig:FCIPCA}(a)] yields an ideal FCI spectrum with threefold degenerate ground states. The $M=1$ truncated form factor [Fig.~\ref{sifig:FCIPCA}(b)] produces less degenerate, but still FCI-like, ground states. The $M=20$ truncated form factor [Fig.~\ref{sifig:FCIPCA}(c)] recovers a spectrum similar to the original. Notably, PCA using only FCI form factors approximates the data around the mean of the FCI set, \AKW{$\bar{\mathbf{x}}$}, which remains an FCI form factor.

\begin{figure}[h!]
\begin{center}
\setlabel{pos=nw,fontsize=\large,labelbox=false}
\xincludegraphics[scale=0.27,label=a]{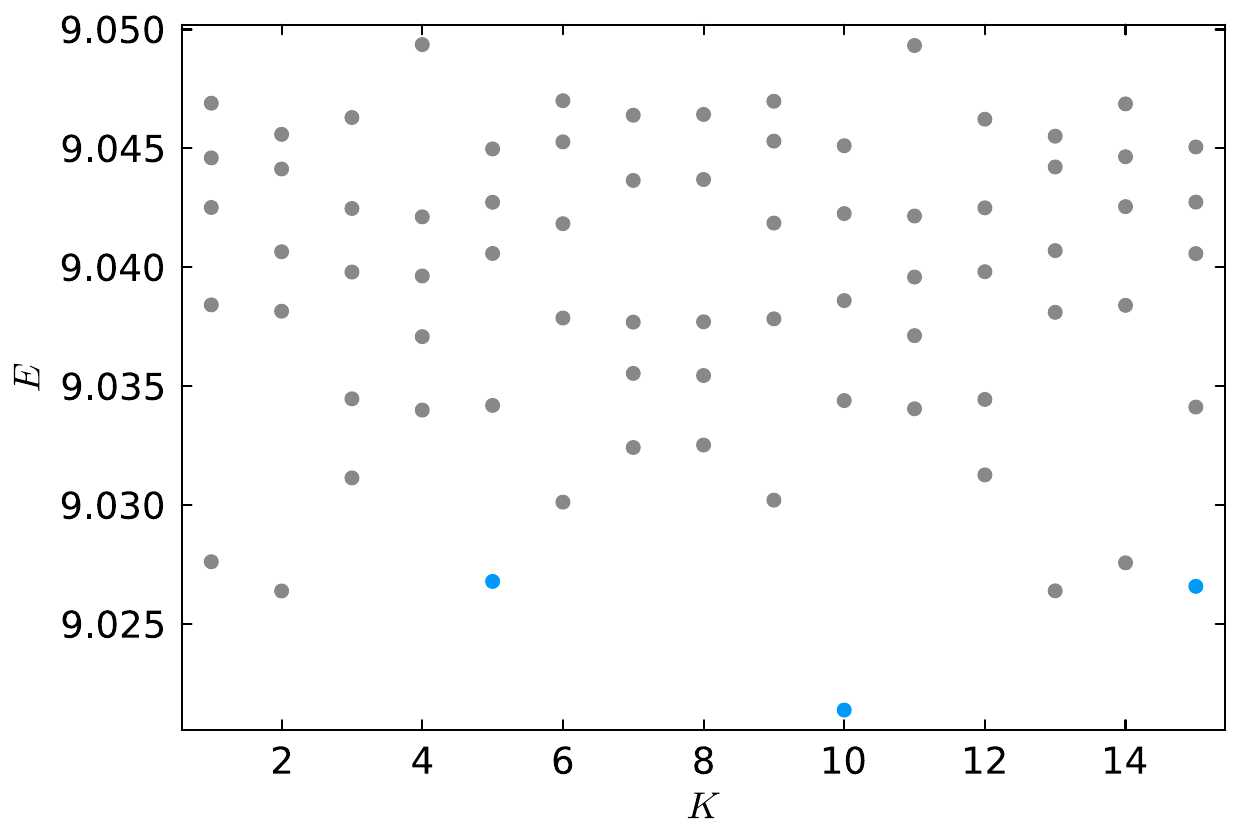}
\xincludegraphics[scale=0.27,label=b]{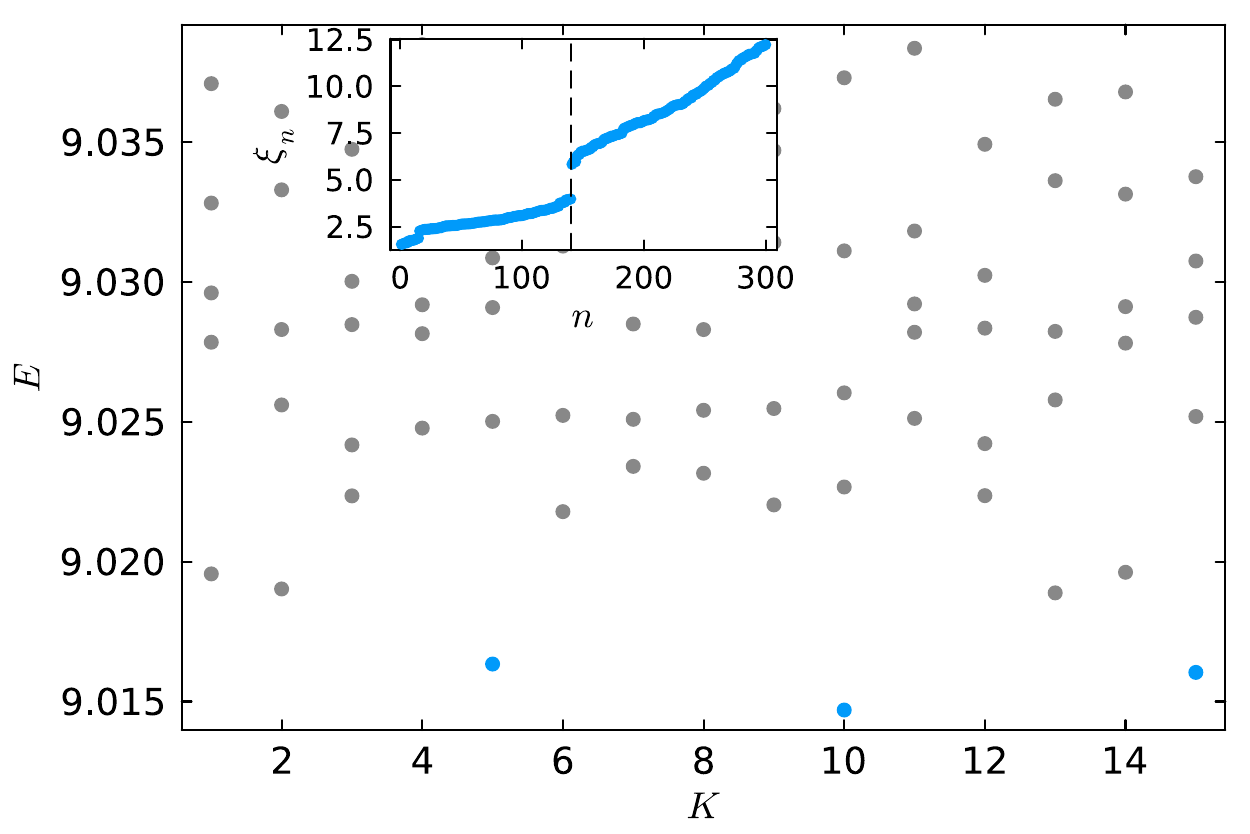}
\xincludegraphics[scale=0.27,label=c]{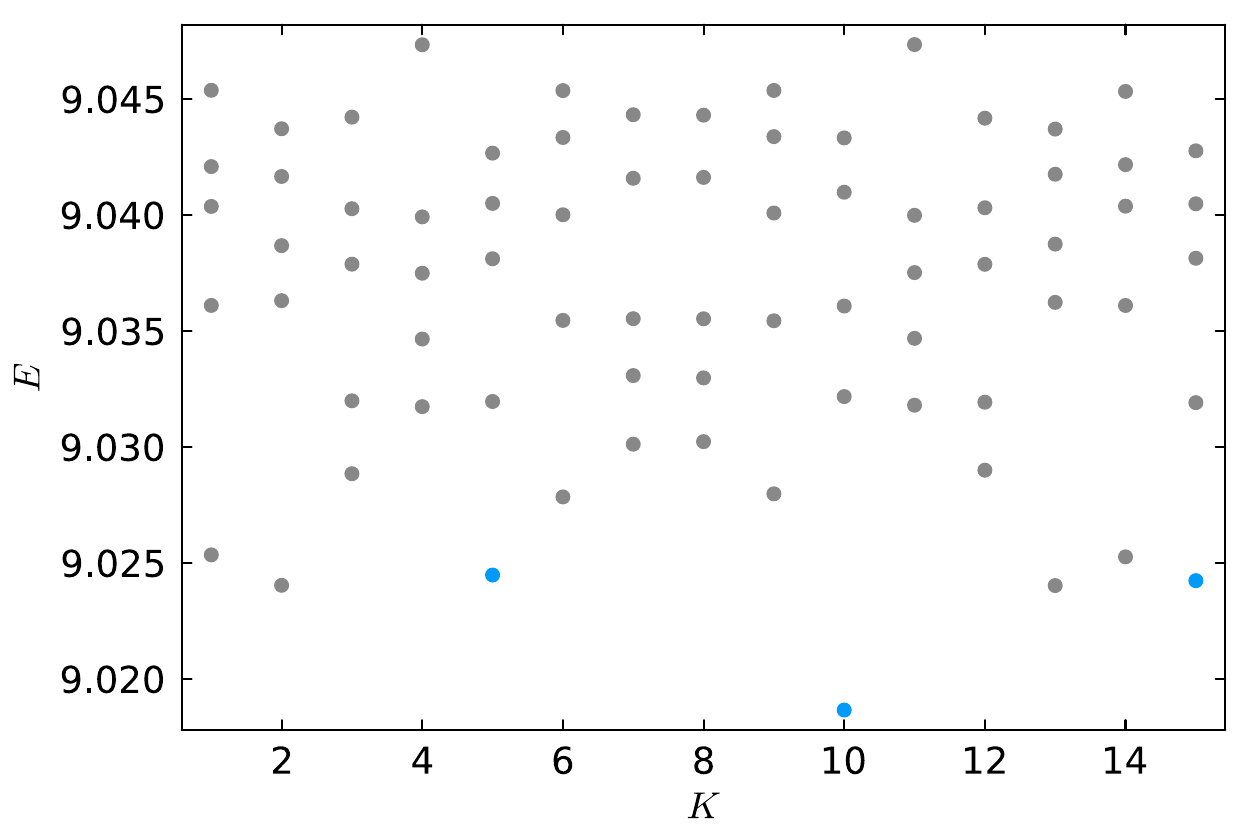}
\caption{\label{sifig:nonFCIPCA}
Principal-component truncation modifies the many-body spectrum of a non-fractional-Chern-insulator form factor. The figure shows exact-diagonalization (ED) spectra for principal component analysis (PCA) truncations of a non-fractional-Chern-insulator (non-FCI) form factor at $\tilde{\alpha}=2.69$ with $N_x=3$ and $N_y=5$. (a) Original form factor. (b) PCA-truncated form factor with $M=10$ retained principal components. (c) PCA-truncated form factor with $M=20$ retained principal components.
}
\end{center}
\end{figure}

We also test whether PCA trained on FCI form factors can approximate previously unseen non-FCI form factors (not included in the PCA training set). In Fig.~\ref{sifig:nonFCIPCA}, we show results for non-FCI form factors at $\tilde{\alpha}=2.69$ with $c_0=0$. The original form factor [Fig.~\ref{sifig:nonFCIPCA}(a)] yields a non-FCI spectrum with no gap and no ground state degeneracy (other than the gap due to finite size effect). The $M=10$ truncated form factor [Fig.~\ref{sifig:nonFCIPCA}(b)] unexpectedly produces a gapped spectrum with three-fold quasi-degenerate ground states, resembling an FCI. The $M=20$ truncated form factor [Fig.~\ref{sifig:nonFCIPCA}(c)] stabilizes a non-FCI spectrum similar to the original one. These results indicate that PCA trained only over FCI data can approximate and extrapolate testing non-FCI form factors.

\bibliography{reference.bib}